\documentclass[lettersize,journal]{IEEEtran}
\usepackage{amsmath,amsfonts}
\usepackage{algorithmic}
\usepackage{algorithm}
\usepackage{array}
\usepackage[caption=false,font=normalsize,labelfont=sf,textfont=sf]{subfig}
\usepackage{textcomp}
\usepackage{stfloats}
\usepackage{url}
\usepackage{verbatim}
\usepackage{graphicx}
\usepackage{cite}
\usepackage{amssymb}
\usepackage{multirow}
\begin{document}

\title{The Codecfake Dataset and Countermeasures for the Universally Detection of Deepfake Audio}

\author{Yuankun Xie, \emph{Student Member, IEEE,}
	Yi Lu, \emph{Student Member, IEEE,}
	Ruibo Fu, \emph{Member, IEEE,}
	Zhengqi Wen, \emph{Member, IEEE,}
	Zhiyong Wang, \emph{Student Member, IEEE,}
	Jianhua Tao, \emph{Senior Member, IEEE,}
	Xin Qi, \emph{Student Member, IEEE,}
	Xiaopeng Wang,  \emph{Student Member, IEEE,}
	Yukun Liu, \emph{Member, IEEE,}
	Haonan Cheng, \emph{Member, IEEE,}
	Long Ye, \emph{Member, IEEE,}
	Yi Sun, \emph{Member, IEEE}

\thanks{Yuankun Xie, Haonan Cheng, Long Ye are with State Key Laboratory of Media Convergence and Communication, Long Ye is with School of Data Science and Intelligent Media, Communication University of China, Beijing, China. (e-mail: xieyuankun@cuc.edu.cn.)}
\thanks{Ruibo Fu is with Institute of Automation, Chinese Academy of Sciences, Beijing, China.}
\thanks{Yi Lu, Zhiyong Wang, Xin Qi, Xiaopeng Wang, Yukun Liu are with School of Artificial Intelligence, University of Chinese Academy of Sciences, Beijing, China.}
\thanks{Jianhua Tao is with Department of Automation, Tsinghua University, Beijing, China. Jianhua Tao, Zhengqi Wen are with Beijing National Research Center for Information Science and Technology, Tsinghua University, Beijing, China.}
\thanks{Yi Sun is with School of Cyberspace science and Technology, Beijing Institute of Technology, Beijing, China.}
\thanks{Yuankun Xie and Yi Lu are co-first authors. Jianhua Tao, Zhengqi Wen and Ruibo Fu are corresponding authors. (e-mail: ruibo.fu@nlpr.ia.ac.cn.)}}

\markboth{Journal of \LaTeX\ Class Files,~Vol.~XX, No.~XX, December~XXXX}%
{Shell \MakeLowercase{\textit{et al.}}: A Sample Article Using IEEEtran.cls for IEEE Journals}

\maketitle

\begin{abstract}
With the proliferation of Audio Language Model (ALM) based deepfake audio, there is an urgent need for generalized detection methods. ALM-based deepfake audio currently exhibits widespread, high deception, and type versatility, posing a significant challenge to current audio deepfake detection (ADD) models trained solely on vocoded data. To effectively detect ALM-based deepfake audio, we focus on the mechanism of the ALM-based audio generation method, the conversion from neural codec to waveform. We initially constructed the Codecfake dataset, an open-source, large-scale collection comprising over 1 million audio samples in both English and Chinese, focus on ALM-based audio detection.    
As countermeasure, to achieve universal detection of deepfake audio and tackle domain ascent bias issue of original sharpness aware minimization (SAM), we propose the CSAM strategy to learn a domain balanced and generalized minima. 
In our experiments, we first demonstrate that ADD model training with the Codecfake dataset can effectively detects ALM-based audio. Furthermore, our proposed generalization countermeasure yields the lowest average equal error rate (EER) of 0.616\% across all test conditions compared to baseline models. The dataset\footnote{https://zenodo.org/records/13838106}and associated code\footnote{https://github.com/xieyuankun/Codecfake}are available online.
\end{abstract}


\begin{figure}[!t]
	\centering
	\includegraphics[width= 2.5in]{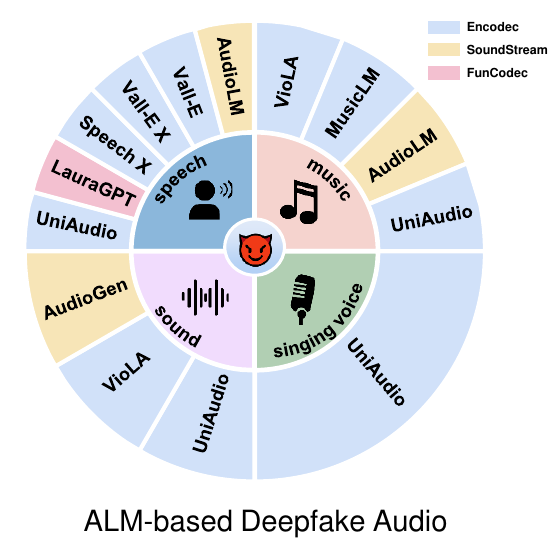}
	\hfil
	\caption{ALM-based deepfake audio. The inner circle represents different audio types, while the outer circle represents different ALM-based generation methods, with different colors indicating the use of different codec methods.}
	\label{fig:ALM}
\end{figure}

\section{Introduction}
With the development of Audio Language Model (ALM) based audio generation methods, an increasing number of deepfake audios are being produced. 
These ALM-based deepfake audio poses significant challenges to current audio deepfake detection (ADD) methods: \begin{itemize}
\item{\textbf{Widespread.} Compared to the previous generation paradigm that relied on acoustic models and back-end vocoders, the novel ALMs employs a neural codec-based generative paradigm. By integrating language models, it significantly lowers the bar for audio generation, allowing anyone to easily create synthetic content. As a result, ALM deepfake audio is becoming widespread, posing a challenge for current ADD models trained solely on vocoded data.}

\item{\textbf{High deception.} ALMs can generate high-quality deepfake audio enhancing its deception. They demonstrate robustness by synthesizing high-quality speech even under low-quality and low-resource conditions, posing challenges for current ADD methods. }

\item{\textbf{Type versatility.} The diversity of audio categories poses challenges for current ADD methods trained by speech. Unlike vocoders, which commonly require specific datasets for training, ALMs leverage a multitude of diverse audio datasets, facilitating the generated audio with versatility type. As a result, ALMs exhibit ability to generate audio beyond the realm of speech, encompassing sound effects, singing voices, and music\cite{yang2023uniaudio}. } 
\end{itemize}

In light of these challenge, the necessity for developing methods of detecting the deepfake audio become increasingly apparent. Currently, researchers have undertaken a series of studies on ADD around competitions such as ASVspoof series \cite{nautsch2021asvspoof,liu2023asvspoof} and ADD Challenge series \cite{yi2022add}. Advanced methods have been proposed that can achieve an intra-domain Equal Error Rate (EER) of less than 1\% \cite{eom22_interspeech, martin2022vicomtech}. However, these methods often experience performance degradation when faced with cross-domain scenarios \cite{muller22_interspeech}. This can be attributed to the inability of ADD models to generalize and detect unseen spoofing methods that were not present in the training set. 

To address the performance degradation of ADD models in the presence of unknown forgery methods, the most direct approach is data augmentation, which aims to expand the training distribution for improving generalizability. Especially in the field of ADD, training ADD models with novel deepfake methods can significantly enhance performance. For traditional audio deepfake generation methods, a vocoder is the core in the generation backend. Therefore, it is significant to augment the data with various vocoders to effectively detect vocoder-based deepfake audio. Recently, Wang et al. \cite{wang2023can} employed multiple neural vocoders to create a large-scale dataset of vocoded data and demonstrated that this approach can improve generalizability in cross-domain testing. Wu et al. \cite{wu2022ustc} also utilized several vocoders to synthesize fake audio based on genuine audio from the training set, achieving the best performance in ADD2023 Track1.2.

\begin{figure*}[!t]
	\centering
	\includegraphics[width= 6.5in]{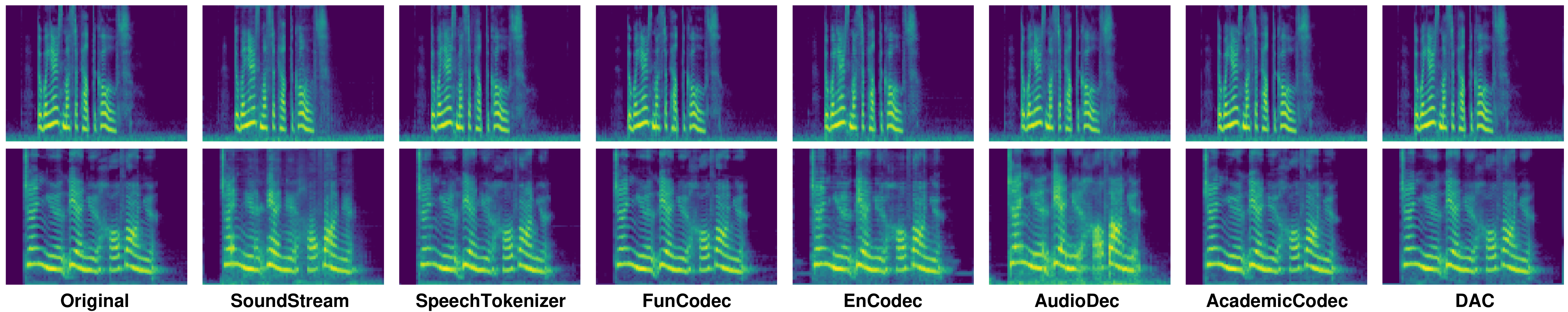}
	\hfil
	\caption{Mel-spectrogram of the original audio alongside 7 codec-based audio samples generated from the original. The top row is generated from VCTK, and the bottom row is generated from AISHELL3.}
	\label{fig:mel}
\end{figure*}

The above studies have explored the core component, vocoders, of deepfake audio, and augmented them based on real audio to better detect vocoder-based deepfake audio. However, current mainstream ALM-based deepfake audio does not utilize vocoder for waveform generation; instead, they employ neural codec approaches to generate audio from discrete codec. As illustrated in Figure \ref{fig:ALM}, existing ALM universally adopt different neural codecs to construct discrete encodings. These ALMs based on codecs often facilitate various audio generation tasks and demonstrate robust generalization capabilities on unseen tasks \cite{yang2023uniaudio}. Furthermore, there exhibit only subtle differences between the features of the original and codec-based audio as shown in Figure \ref{fig:mel}. This poses a significant challenge for ADD methods. Consequently, there is an urgent need to construct a neural codec dataset based on real audio and develop a generalized countermeasure to accurately detect ALM-based deepfake audio.

In this paper, we introduce Codecfake, a initial large-scale dataset for detecting ALM-based deepfake audio, comprising over 1 million audio samples in two languages: English and Chinese. Specifically, we select 7 representative open-source neural codec models to construct fake audio. These codec types encompass the current mainstream ALM-based audio generation models.  To effectively assess the capability of ADD models trained with Codecfake, we set up diverse test conditions, encompassing unseen codec-based audio, codec-based audio synthesized with varying parameters, and in-the-wild ALM-based audio. For countermeasure, we propose a generalized training strategy called co-training sharpness aware minimization (CSAM) to effective detect both the vocoder-based audio and codec-based audio. Experiments demonstrate that vocoder-trained ADD models are insufficient for effectively detecting codec-based audio. Leveraging the proposed Codecfake dataset and CSAM strategy for training, we achieve the lowest 0.616\% average EER on diverse test condition.

\section{Related Work}
This section presents a brief introduction to the development of vocoder-based deepfake approaches, codec-based deepfake approaches and audio deepfake detection datasets.

\subsection{Vocoder-based deepfake audio}
Neural TTS systems primarily consist of three main components: a text analysis module, an acoustic model, and a vocoder. Deepfake audio detection mainly focuses on the latter two aspects.
In terms of the acoustic model, Tacotron2 \cite{shen2018natural} and Transformer-TTS \cite{li2019neural} are capable of converting input text into Mel spectrograms frame by frame, producing highly natural speech. However, they are autoregressive in nature, resulting in slower generation speeds. To address the aforementioned issue, FastSpeech1/2 \cite{ren2019fastspeech, ren2020fastspeech} employ external alignment to train an additional module that predicts the audio frames corresponding to text tokens, enabling parallel inference. Nonetheless, in order to learn the number of frames corresponding to phonemes, they require an external model to assist in the training of the duration predictor. Glow-TTS \cite{kim2020glow} is a parallel generation method based on normalizing flows. It introduces the Monotonic Alignment Search algorithm, which addresses the aforementioned issue. Recently, diffusion model-based methods have gradually gained momentum in the field of speech synthesis, owing to their capacity to synthesize high-quality audio. This trend is exemplified by models such as Grad-TTS \cite{popov2021grad} and DiffSinger \cite{liu2022diffsinger}.

Since the publication of WaveNet \cite{oord2016wavenet}, research on neural vocoders has continued to emerge. WaveNet is capable of generating extremely high-quality audio, but its autoregressive model leads to excessively slow inference speeds, rendering it unsuitable for real-time generation. Parallel WaveNet \cite{oord2018parallel} and WaveGlow \cite{prenger2019waveglow}, based on Normalizing Flows, attempt to accelerate inference while preserving quality. Nowadays, GAN-based vocoders such as MelGAN, Parallel WaveGAN, and HiFiGAN \cite{kumar2019melgan, yamamoto2020parallel, kong2020hifi} are widely utilized due to their advantages of fast inference speed, high quality, and small model size. Recently, vocoders based on diffusion, such as DiffWave and WaveGrad \cite{kong2020diffwave, chen2020wavegrad}, have also been proposed, but they have not been widely adopted yet. To streamline the training process, end-to-end models such as VITS \cite{kim2021conditional}, NaturalSpeech1/2/3 \cite{tan2022naturalspeech, shen2023naturalspeech, ju2024naturalspeech}, etc., integrate the acoustic model and vocoder, leading to end-to-end TTS designs and training. This approach has seen significant development and has become the current SOTA in the field.

In the field of Voice Conversion (VC), the overall architecture of mainstream VC models today is essentially similar to TTS, comprising both an acoustic model and a vocoder. However, the key difference lies in the input to the acoustic model, which is no longer text features but rather specific content features extracted from the audio to be converted. Additionally, speaker characteristics from the target audio are incorporated into the acoustic model. Early content features primarily used Phonetic PosteriorGrams (PPGs) \cite{sun2016phonetic}, whereas contemporary approaches mainly rely on audio pre-training models such as HuBERT \cite{hsu2021hubert} for feature extraction. Speaker characteristics are typically represented by d-vectors \cite{variani2014deep} or intermediate layer features from speaker verification models. In summary, current vocoder-based audio generation methods can produce audio with such high quality and realism that it is often indistinguishable from real audio for people. However, there is still room for improvement in style and diversity.

\subsection{Codec-based deepfake audio}
Language modeling has achieved significant success in the field of natural language processing (NLP). Similarly, audio, as a form of temporal information, encompasses not only content but also rich information such as timbre, emotions, rhythm, acoustic details, among others. This offers great potential for language models to handle audio information effectively. Neural audio codec models were initially introduced into the domains of audio compression and data transmission. The encoder compresses audio into codec codes for transmission, while the decoder at the receiving end decodes the received codec codes to reconstruct the audio. To leverage language models for discrete modeling of audio, it is essential to utilize audio codecs to convert continuous audio into discrete codes. Recently, a plethora of ALMs based on audio codec models have emerged, yielding remarkably excellent synthesis results \cite{borsos2023audiolm, kreuk2022audiogen, wang2023neural, agostinelli2023musiclm, zhang2023speak, wang2023viola, copet2024simple, wang2023speechx, chen2023lauragpt, yang2023uniaudio}.

Traditional audio codecs can be classified into two main categories: waveform codecs and parametric codecs. Waveform codecs make fewer assumptions about the audio content and can produce high-quality generic audio. However, at lower bit rates, they often introduce coding artifacts. Parametric codecs, on the other hand, employ strong priors and reconstruct audio by encoding and decoding parameters. The reconstructed audio perceptually resembles the original audio but is not faithfully reconstructed. An ideal audio codec should strive to faithfully reconstruct audio while also being able to generalize across various audio content, regardless of the specific characteristics of the audio.

Recently, neural audio codecs have demonstrated efficiency in audio compression and decoding, surpassing traditional audio codec models. SoundStream \cite{zeghidour2021soundstream} is one of the earliest universal neural audio codecs, with a single model supporting various different bitrates. Its architecture employs streaming SEANets \cite{roblek2020seanet} as its encoder-decoder, with quantization utilizing Residual Vector Quantization (RVQ) combined with a speech enhancement module. The EnCodec \cite{defossez2022high} architecture bears a striking resemblance to the SoundStream architecture, but it incorporates LSTM layers \cite{graves2012long} and leverages Transformer-based speech models to model RVQ codes, enhancing the model's generation efficiency while improving discriminative loss and enhancing generation performance. Building upon Encodec, AudioDec \cite{wu2023audiodec} introduces improvements by employing a group convolution mechanism to accelerate inference speed for real-time transmission. Additionally, it incorporates a multi-period discriminator from HiFi-GAN to ensure the generation of high-fidelity audio. AcademiCodec \cite{yang2023hifi} introduces group-residual vector quantization and employs multiple parallel RVQ groups to enhance audio reconstruction performance while managing codebook size limitations. Descript-audio-codec (DAC) \cite{kumar2024high} stands out as a versatile neural codec model capable of preserving high-quality audio across a broad frequency spectrum. It incorporates advanced residual vector quantization techniques and utilizes periodic activation functions. SpeechTokenizer \cite{zhang2024speechtokenizer} leverages semantic tokens derived from Hubert L9 \cite{hsu2021hubert} as a guide for the RVQ process, aiding in the separation of content and acoustic information. Content information is processed in the initial layer of the tokenizer, while subsequent layers retain acoustic details. FunCodec \cite{du2023funcodec} operates on audio encoding in the frequency domain, requiring fewer parameters for comparable performance. Moreover, it assesses the influence of semantic cues on speech codecs, thereby enhancing speech quality particularly at low-bitrates.

With the rapid advancement of audio codecs and Large Language Model models, ALM-based audio generation models have emerged successively. AudioLM \cite{borsos2023audiolm} stands as a pioneering model in integrating codec codes for language modeling, employing a hierarchical approach comprising two distinct stages. In the initial stage, semantic tokens are generated utilizing a self-supervised w2v-BERT model \cite{chung2021w2v}. Subsequently, in the second stage, these tokens serve as conditioning elements to generate acoustic tokens employing a SoundStream neural codec. VALL-E, VALL-E X, and SpeechX are neural codec language models designed to generate discrete codes derived from EnCodec, utilizing either textual or acoustic inputs. VALL-E \cite{wang2023neural} is capable of generating high-quality audio with specific timbres using only 3 seconds of unseen speaker's audio. VALL-E X \cite{zhang2023speak} can generate high-quality audio in the target language with just a single utterance of the source language audio as a prompt. SpeechX \cite{wang2023speechx} introduces a unified framework that addresses not only zero-shot TTS but also various types of speech transformation tasks, including speech enhancement and speech editing. VioLA, AudioPaLM, and LauraGPT are capable of generating both audio and text. VioLA \cite{wang2023viola} is a decoder-only generative model that supports tasks such as speech recognition, synthesis, and translation. It employs task ID and language ID along with text tokens and audio tokens for modeling, with audio tokens extracted by Encodec. AudioPaLM \cite{rubenstein2023audiopalm} constructs a unified vocabulary containing both text and audio tokens. It is a decoder-only autoregressive model capable of processing and generating both text and speech. The audio tokenization method of AudioPaLM resembles that of AudioLM. Additionally, AudioPaLM adopts the SoundStream model and extends it to SoundStorm. LauraGPT \cite{chen2023lauragpt} is a versatile language model based on the decoder-only text-based language model Qwen-2B \cite{bai2023qwen}. It can handle both audio and text inputs and generate outputs in either mode. LauraGPT utilizes a Conformer \cite{gulati2020conformer} encoder to encode input audio into continuous representations and employs FunCodec discrete codes to decode the output audio. UniAudio \cite{yang2023uniaudio} employs language models to accomplish various audio generation tasks, including speech generation, audio generation, music generation, singing generation, denoising, and more, by inputting text. It utilizes a large-scale full Transformer to predict the first layer of codec codes and a smaller local Transformer to predict subsequent codec layers, thereby enhancing prediction speed. The encoder-decoder model in UniAudio is modified from EnCodec. UniAudio has achieved or even surpassed current mainstream traditional generative models in multiple domains of audio generation tasks.

\begin{table*}[t]
	\caption{Characteristics of representative datasets in the field of ADD.}
	\label{tab:dataset}
	\centering
	\setlength{\tabcolsep}{2mm} 
	\renewcommand{\arraystretch}{1.2}
	\begin{tabular}{|c|c|c|c|c|c|c|c|c|c|c|c|c|}
		\hline
		Dataset &Year &Languages &Methods  &Back-end &Type&Real  &Fake &Total\\
		\hline
		ASVspoof2019LA \cite{nautsch2021asvspoof}&2019 &English &19 &Vocoder &Speech &10,256 &90,192 &100,448\\
		\hline
		ASVspoof2021LA \cite{liu2023asvspoof} &2021 &English &19 &Vocoder &Speech &14,816 &133,360 &148,176\\ 
		\hline
		ASVspoof2021DF \cite{liu2023asvspoof} &2021 &English &100+ &Vocoder &Speech &14,869 &519,059 &533,928\\
		\hline
		WaveFake \cite{frank2021wavefake}&2021 &English/Japanese &7 &Vocoder &Speech &0 &117,985&117,985 \\
		\hline
		ITW \cite{muller22_interspeech} &2022 &English &- &- &Speech &19,963 &11,816 &31,779 \\
		\hline
		TIMIT-TTS \cite{salvi2023timit} &2022 &English &12 &Vocoder &Speech &0 &5,160&5,160\\
		\hline
		ADD2022T1.2 \cite{yi2022add} &2022 &Chinese &- &Vocoder &Speech &36,953 &123,932&160,885\\
		\hline
		CFAD \cite{ma2022fad}    &2023 &Chinese &12 &Vocoder &Speech &38,600 &77,200&115,800\\		
		\hline
		ADD2023T1.2 \cite{yi2023add} &2023 &Chinese &- &Vocoder &Speech &172,819 &113,042&285,861\\	
		\hline
		MLAAD \cite{muller2024mlaad} &2024 &Multilanguages &54 &Vocoder &Speech &0 &76,000&76,000\\
		\hline
		ASVspoof5 \cite{wang2024asvspoof} &2024 &English &32 &Vocoder &Speech &188,819 &815,262 &1,004,081\\		
		\hline
		Codecfake &2024 &English/Chinese &7 &Neural Codec &Speech/Sound &132,277 &925,939&1,058,216\\		
		\hline
	\end{tabular}
	
\end{table*}

\subsection{Audio Deepfake Detection Dataset}
With the fast propagation and high-quality of deepfake audio, the filed of ADD has attracted increasing attention from endeavors. The first step in expanding a field of research is to establish corresponding datasets. In the current task of ADD, numerous dataset initiatives are propelling the development of the field. In the Table \ref{tab:dataset}, 

The two most prominent dataset initiatives are the ASVspoof Challenge series  \cite{nautsch2021asvspoof,liu2023asvspoof} and the ADD Challenge series \cite{yi2022add,yi2023add}.

Surrounding the ASVspoof Challenge series, especially in ASVspoof2019LA (19LA)\cite{nautsch2021asvspoof}, numerous outstanding works on ADD have emerged. The training set of 19LA includes six vocoder-based synthesized TTS and VC utterances, labeled as A01-A06. The test set of 19LA comprises unseen spoofing methods A07-A19. This training and testing protocol aims to select generalizable anti-spoofing models. In subsequent challenges like ASVspoof2021LA (21LA) and ASVspoof2021DF (21DF) track \cite{liu2023asvspoof}, participants are still limited to using the training set from 19LA. However, the test sets introduce more challenging conditions. Specifically, 21LA track includes 19LA evaluation set condition from different real telephone systems, while the 21DF track involves more than hundreds of vocoder-based audio samples generated from various source domains and under various compression conditions. Currently, ASVspoof5 \cite{wang2024asvspoof} has been successfully held, focusing on crowdsourced datasets and including adversarial attacks and various compression conditions.

Regarding ADD Challenge series, the dataset not only includes a binary classification between real and fake but also encompasses additional tasks such as partially deepfake detection and source tracking. The binary classification task primarily focuses on low-quality Chinese ADD task, which means that the utterances take into account various real-world noises and background music effects.

In addition to the two challenge series datasets mentioned, there are also other ADD datasets designed to address various issues. WaveFake \cite{frank2021wavefake} presents a vocoder-based audio deepfake dataset collected from ten sample sets across six different network architectures, spanning two languages. In the wild (ITW) dataset \cite{muller22_interspeech} collects both real and fake audio samples from publicly available sources like social networks and video streaming platforms. It potentially contains background noises and is designed to assess the generalizability of ADD models. CFAD \cite{ma2022fad} is the first Mandarin audio spoofed dataset consisting of twelve different vocoder-based generation algorithms including both TTS and VC types. MLAAD \cite{muller2024mlaad} is a recent work that covers 23 languages and up to 54 TTS and VC spoofing methods.

Although current ADD datasets offer a rich variety of spoof types, most of their generation back-ends are based on vocoders. With the advancements in ALM generation method, there is an urgent need for datasets generated based on codecs to facilitate research into detecting ALM-based audio.

\section{Dataset Design}
\subsection{Overview}
\textbf{Motivation.}
We employ speech codec models to construct a dataset of forged audio, aiming to enhance the detection performance of fake audio detection models against ALM-based audio. 
This choice is motivated by: \begin{itemize}
\item{\textbf{Improving generalizability.} Traditional ADD datasets are primarily constructed using vocoders, while ALM primarily utilizes codec-based methods to decode discrete codes generated by language models into audio, rather than using vocoders. Consequently, models trained on vocoder-based datasets have limited generalization capabilities for detecting ALM-generated fake audio.}
\item{\textbf{Strengthening codec representation learning.} ALM-based audio decoders apply neural codec algorithms to convert discrete codes into waveforms, which can often give rise to artifacts due to the upsampling convolution module\cite{pons2021upsampling}. Training with Codecfake enables ADD models to learn the patterns of codec artifacts, thus enhancing their ability to effectively detect ALM-based deepfake audio.}
\item{\textbf{Practical needs.} From a practical standpoint, most current ALMs are not open-source, and the cost of training them is prohibitive for most individuals. In contrast, neural audio codec models have reliable implementations and are widely utilized in ALM systems.}
\end{itemize}

\textbf{Strength.}
Compared to existing ADD datasets, our proposed Codecfake dataset presents several key advantages: \begin{itemize}
\item{As shown in Table \ref{tab:dataset}, traditional ADD datasets are primarily constructed using vocoders, whereas our proposed Codecfake dataset leverages neural codec-based audio, enabling more effective detection of ALM-based audio.}
\item{The Codecfake dataset provides a wide range of test conditions simulating real-world scenarios, encompassing various audio types (speech and sound) and ALM-based audio (A1, A2).}
\item{Our proposed Codecfake dataset comprises over 1 million samples across two languages and covers codecs used in current ALM implementations.}
\end{itemize}

\textbf{Completeness.}
We conducted a survey of the neural audio codec models predominantly utilized in the current ALM implementations, which primarily include SoundStream, Encodec, and FunCodec, as illustrated in Figure \ref{fig:ALM}. Consequently, our constructed dataset incorporates these three methods among the seven approaches. To ensure the long-term effectiveness of the dataset, we also selected several methods that have not been temporarily used by ALM but have demonstrated good performance and representativeness. This approach enhances the dataset's integrity and future scalability. This will ensure the completeness of the dataset over an extended period.

\subsection{Architectures for generating codec-based fake audio}
In this section, we provide a brief overview of the architectures and differences of the seven neural audio codecs models (F01-F07) used to construct our dataset. Those seven methods encompass all major neural audio codecs models proposed in recent years, ensuring coverage of the codecs types employed by mainstream ALMs and potentially superior neural audio codecs models that may be adopted by future ALMs.

\textbf{F01-SoundStream} \cite{zeghidour2021soundstream}. SoundStream is a milestone achievement in neural audio codecs, employing a classic architecture consisting of encoder, quantizer, and decoder components. Its quantizer utilizes residual vector quantizers (RVQ) \cite{van2017neural}, and it integrates denoising functionality throughout the entire architecture.

\textbf{F02-SpeechTokenizer} \cite{zhang2024speechtokenizer}. SpeechTokenizer utilizes semantic tokens from Hubert L9 as a teacher for the RVQ process. This aids in decoupling content information from acoustic information, with content information processed in the first layer of the tokenizer, while acoustic information is retained in subsequent layers.

\textbf{F03-FunCodec} \cite{du2023funcodec}. FunCodec encodes audio in the frequency domain, requiring fewer parameters to achieve the same effect. Additionally, it evaluates the impact of semantic information on speech codecs, thereby enhancing speech quality at low bit rates.

\textbf{F04-EnCodec} \cite{defossez2022high}. The structure of Encodec is similar to that of SoundStream. It integrates additional LSTM \cite{hochreiter1997long} layers and lightweight Transformer-based language models to model RVQ encoding, thereby accelerating inference speed while maintaining quality.

\textbf{F05-AudioDec} \cite{wu2023audiodec}. Building upon Encodec, AudioDec introduces improvements by employing a group convolution mechanism to accelerate inference speed for real-time transmission. Additionally, it incorporates a multi-period discriminator from HiFi-GAN \cite{su2020hifi} to ensure the generation of high-fidelity audio.

\textbf{F06-AcademicCodec} \cite{yang2023hifi}. AcademiCodec introduces group-residual vector quantization and utilizes multiple parallel RVQ groups, aiming to enhance audio reconstruction performance within the constraints of codebook size.

\textbf{F07-Descript-audio-codec (DAC)} \cite{kumar2024high}. DAC is a versatile neural codec model capable of maintaining high-quality audio across a wide frequency range. It employs enhanced residual vector quantization and utilizes periodic activation functions 	\cite{ziyin2020neural}.
\begin{figure*}[!t]
	\centering
	\subfloat{\includegraphics[width=7in]{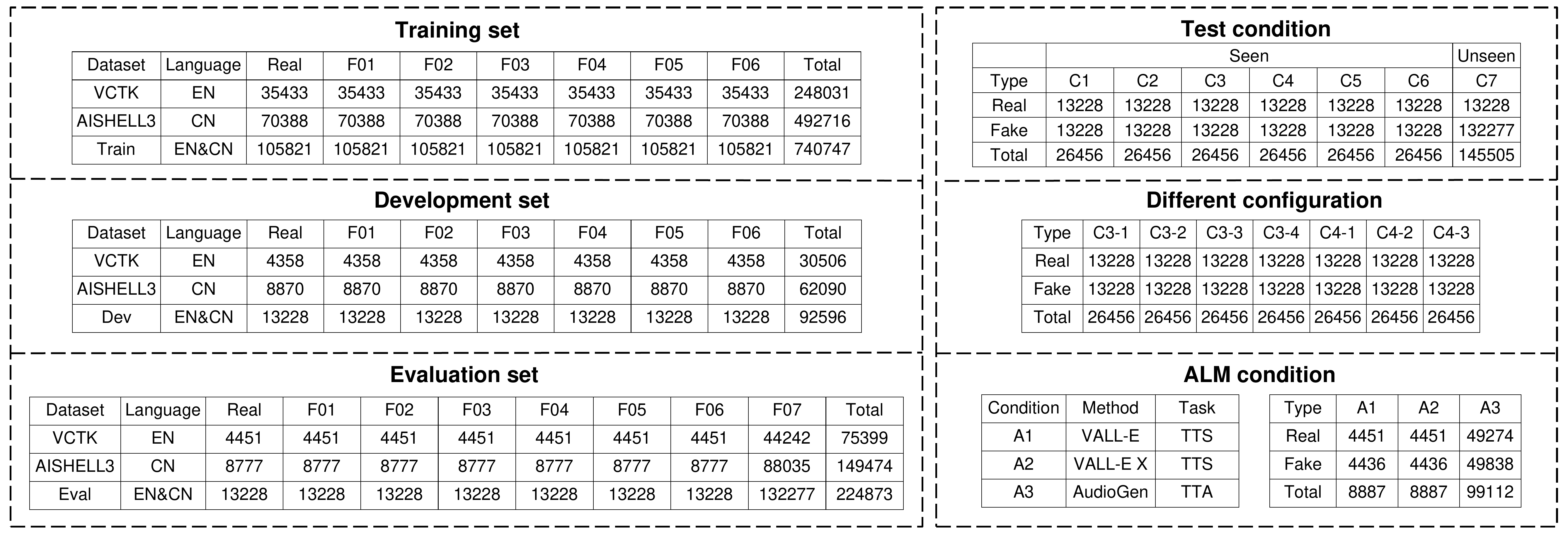}}
	\hfil
	\caption{Partitions and construction of Codecfake dataset. Left part displays the training set, development set, and evaluation set of Codecfake dataset. The right part illustrates the diverse testing conditions of Codecfake.}
	\label{dataset_design}
\end{figure*}

\begin{table}[t]
	\centering	
	\caption{Neural codec models detail.}
	\label{tab:encodec}
	\renewcommand{\arraystretch}{1.2}
	\begin{tabular}{|c|c|c|c|c|}
		\hline
		\textbf{Type} &\textbf{Codec}& \textbf{SR} & \textbf{BPS} & \textbf{Quantizers} \\ \hline
		F01&SoundStream    & 16k & 4k & 8                                            \\ \hline
		F02&SpeechToknizer & 16k & 4k & 8                                           \\ \hline
		F03&FuncCodec      & 16k & 16k & 32                                        \\ \hline
		F04&EnCodec        & 24k & 6k & 8                                           \\ \hline
		F05&AudioDec       & 24k & 6.4k & 8                                        \\ \hline
		F06&AcademicCodec  & 24k & 3k & 4                                           \\ \hline
		F07&DAC            & 44k & 8k & 9                                        \\ \hline
		
	\end{tabular}
\end{table}
\begin{table}[t]
	\centering	
	\caption{Different parameters setting for test condition C3 and C4.}
	\label{tab:different}
	\renewcommand{\arraystretch}{1.2}
	\begin{tabular}{|c|c|c|c|c|}
		\hline
		\textbf{Type} &\textbf{Codec}& \textbf{SR} & \textbf{BPS} & \textbf{Quantizers} \\ \hline
		C3-1&FuncCodec    & 16k & 3k & 12                                            \\ \hline
		C3-2&FuncCodec    & 16k & 6k & 24                                           \\ \hline
		C3-3&FuncCodec    & 16k & 12k & 32                                        \\ \hline
		C3-4&FuncCodec    & 16k & 24k & 32                                           \\ \hline
		C4-1&EnCodec      & 24k & 3k & 4                                       \\ \hline
		C4-2&EnCodec      & 24k & 12k & 16                                           \\ \hline
		C4-3&EnCodec      & 24k & 24k & 32                                        \\ \hline
		
	\end{tabular}
\end{table}

\subsection{The Generation Process of codec-based fake audio}
In this section, we will introduce the training details of neural audio codecs, the datasets used, as well as the inference process and the construction of the Codecfake dataset.

During the training phase, neural codec models are trained using the LibriTTS \cite{zen2019libritts} dataset, which is a multi-speaker English corpus comprising approximately 585 hours of read English speech at a sampling rate of 24kHz. The LibriTTS corpus is specifically designed for research in text-to-speech systems and is derived from the original materials of the LibriSpeech \cite{panayotov2015librispeech} corpus. We typically set the parameters of each model to the default values mentioned in the original paper or to parameter settings widely adopted by ALMs for training, continuing training until the model converges.

During the inference phase, we utilize the trained seven neural codec models to re-encode and decode data from the VCTK \cite{Veaux2017CSTRVC} (This CSTR VCTK Corpus includes speech data uttered by 110 English speakers with various accents.) and AISHELL3 \cite{shi21c_interspeech} (The corpus contains roughly 85 hours of emotion-neutral recordings spoken by 218 native Chinese mandarin speakers) datasets separately. More neural codec models details is ilustrated in Table \ref{tab:encodec}.

\subsection{Overall Statistics}
Our proposed Codecfake dataset consists of 1,058,216 audio samples, including 132,277 real samples (44,242 samples from VCTK and 88,035 samples from AISHELL3) and 925,939 fake samples generated by seven different codec methods. Regarding ADD experiments, we divided the 132,277 source real samples into a training subset of 105,821 samples, a development subset of 13,228 samples, and an evaluation subset of 13,228 samples. The same dataset splitting strategy was applied to the fake audio generated using the six different forgery methods (F01-F06). We designated F07 as an unseen fake methods for the generalization verification of the ADD model.  This implies that there are no F07 utterances in either the training or development sets. The training set of Codecfake comprises a total of 740,747 samples, while the development set contains 92,596 samples, and the evaluation set contains 224,873 samples. Thus, in total, Codecfake dataset consists of 1,058,216 audio samples. More dataset details can seen in the left part of Figure \ref{dataset_design}. 

\subsection{Testing Condition}
To effectively verify whether ADD models trained with Codecfake can accurately detect codec-based audio, and even ALM-based audio, we created diverse test conditions, as illustrated on the right side of the Figure \ref{dataset_design}. For evaluation, we established seven conditions (C1-C7) for codec methods F01-F07, with each condition including the evaluation subset of real audio. 
It is worth mentioning that condition C7 represents unseen data for the training set, thus its fake part includes a total of 132,277 samples. 

To further validate whether codec-based audio generated by different parameters of the audio codec model will have a significant impact on the accuracy of the ADD model, we synthesized different parameters for conditions C3 and C4. Specifically, for C3, we selected four different configurations labeled as C3-1 to C3-4, and for C4, we selected three configurations labeled as C4-1 to C4-3. These configurations are options for training parameters in F03 and F04. 

To verify whether codec-trained ADD model can detect audio generated by ALM, we created ALM conditions A1-A3. The generation of these three tasks adopts their open-source default configurations. Among them, A1 and A2 represent text-to-speech (TTS) task, while A3 represents a text-to-audio (TTA) task. A1 is generated by VALL-E \cite{wang2023neural}, A2 by VALL-E X \cite{zhang2023speak}, and A3 by AudioGen \cite{kreuk2022audiogen}. For A1 and A2, we select the text sourced from the VCTK evaluation set corresponding to 4451 text items. However, the speaker p315 does not have corresponding text, so the generated audio for A1 and A2 consists of only 4437 audio samples. For the real source domain of A3, we utilize the training subset of the Audiocaps \cite{kim2019audiocaps}, which was employed during the training phase of AudioGen, generating 49,838 fake audios based on its corresponding texts. Given that some of the original Audiocaps audio links are no longer active, we use the currently available 49,274 audio samples as the real source domain for A3.

\section{Audio Deepfake Detection}
\subsection{Baseline model}
\begin{figure}[!t]
	\centering
	\includegraphics[width= 3in]{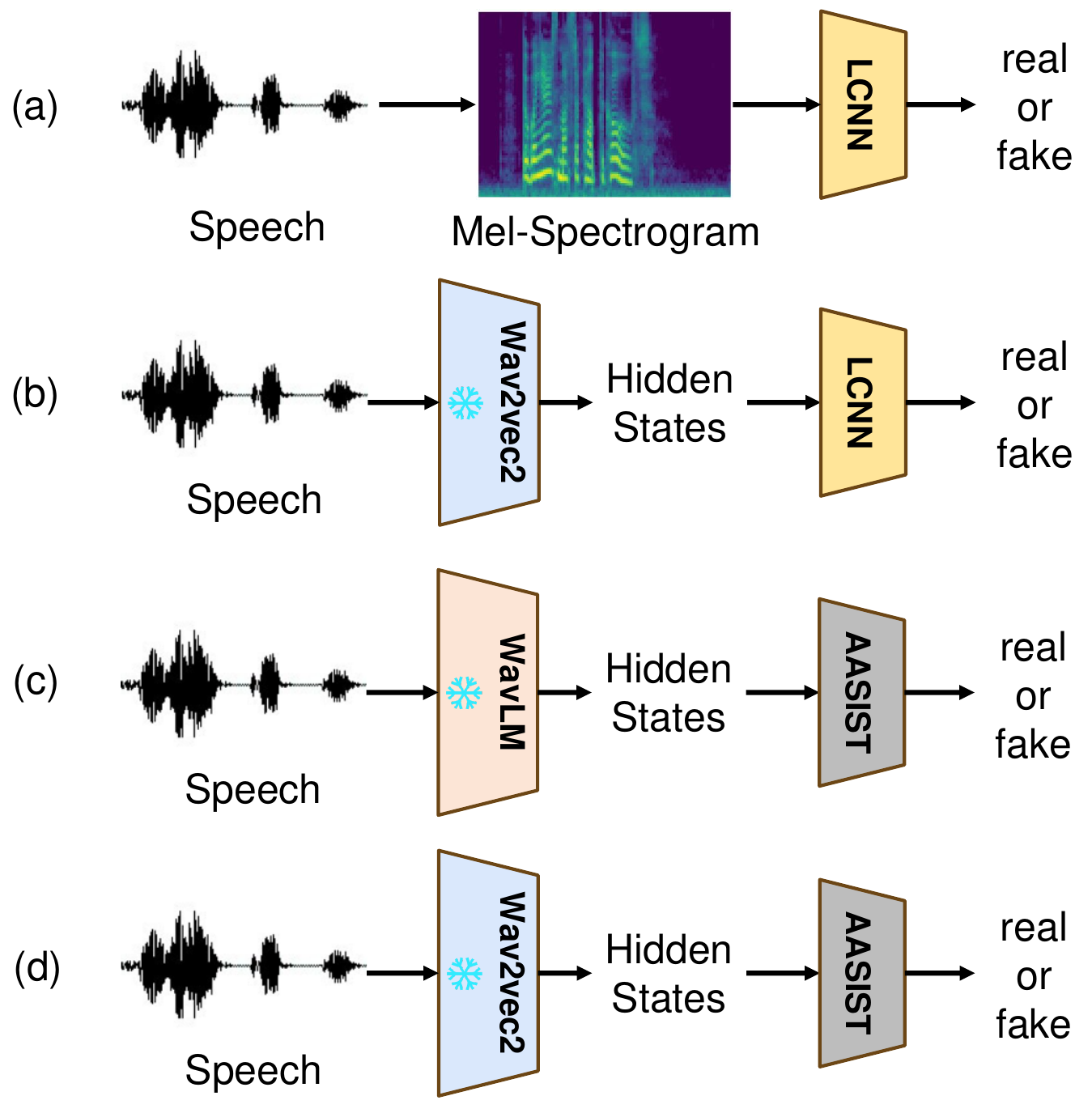}
	\hfil
	\caption{Four baseline ADD models for evaluation. (a), (b), (c), (d) represents Mel-LCNN, W2V2-LCNN, WavLM-AASIST, W2V2-AASIST, respectively.}
	\label{fig:backbone}
\end{figure}

For ADD model, we select four baseline models for evaluation as shown in Figure \ref{fig:backbone}. 
For the front-end features, we chose the commonly used handcrafted feature in the field of audio deepfake generation, the Mel-spectrogram, to facilitate the analysis of different codec-based audios. On the other hand, we utilized the frozen pre-trained hidden representation of Wav2vec2-XLS-R (W2V2) \cite{babu2021xls} and WavLM \cite{chen2022wavlm} as input features. For W2V2, existing research has convincingly demonstrated its superior performance in the ADD task \cite{pascu24_interspeech, phukan2024heterogeneity, zhang2024audio}. Besides W2V2, we also adopt WavLM as the front-end feature due to its exceptional performance in other downstream tasks \cite{yang21c_interspeech}. It is worth mentioning that we utilize the fifth hidden state from the frozen pre-trained feature as the final feature, in reference to previous research on frozen hidden states \cite{lee22q_interspeech}, which demonstrated the best performance.

For the backbone model, we employ two popular backbone networks in the field of ADD, LCNN \cite{lavrentyeva2019stc} and AASIST \cite{jung2022aasist}. LCNN is a lightweight convolution-based backbone featuring an MFM layer designed to effectively filter feature channels conducive to authentication. AASIST represents the state-of-the-art backbone in the ADD domain, introducing a novel heterogeneous stacking graph attention layer that models artifacts across heterogeneous temporal and spectral domains using a heterogeneous attention mechanism and stacked nodes. Specifically, we employ the backbone of this version of AASIST \cite{tak2022automatic}, which utilizes a fully connected layer to down-sample the hidden states of W2V2 from 1024 dimensions to 128, matching the dimension of the front-end of AASIST.

\subsection{Countermeasure for generalized ADD method}
To universally detect both codec-based audio and vocoder-based audio, we conducted co-training with 19LA and Codecfake. However, we found that the testing performance of the co-trained model fell between that of vocoder-trained ADD models and codec-trained ADD models. Motivated by \cite{shim23c_interspeech}, we employ sharpness aware minimization (SAM) method, which offers a promising alternative to empirical risk minimization (ERM) \cite{vapnik1991principles} for seeking a flat and generalizable minima in co-training senario. At first, we can define $\theta$ as the weight parameters of a neural network $f$. $\mathcal{D}=\left\{\mathcal{D}_{i}\right\}_{i}^{S}$ denotes the whole dataset $D$ conclude $S$ training domain. Suppose there are $n$ training samples in each source training domain $\mathcal{D}_{i}$. For each audio input $x$ with an binary output label $y$, we can define a co-training ERM optimization as shown in Equation \ref{equation:loss}.
\begin{equation}
	\mathcal{L}(\theta ; \mathcal{D}) = \frac{1}{S n} \sum_{i=1}^{S} \sum_{j=1}^{n} \ell \left(f\left(x_{j}^{i} ; \theta\right), y_{j}^{i}\right),
	\label{equation:loss}
\end{equation}
where $l$ denotes the cross-entropy loss. Training the network typically involves solving a non-convex optimization problem, with the objective of finding an optimal weight vector $\hat{\theta}$ that minimizes the empirical risk $L(\theta; D)$. However, in practice, only optimize ERM tends to overfit to the training set and usually converge to a sharp minima, which lead to a bad generalizability to unseen domain. Thus, SAM is proposed to find a flatter region around the minimum with the following equation:
\begin{equation}
\min _{\theta} \max _{\epsilon:\|\epsilon\| \leq \rho} \mathcal{L}(\theta+\epsilon ; \mathcal{D}).
\label{equation:sam1}
\end{equation}
For a given $\theta$, the maximization in Equation 2 seeks the weight perturbation $\epsilon$ within the euclidean ball of radius $\rho$ that maximizes the empirical loss. However, the maximization in Equation \ref{equation:sam1} is costly. In SAM algorithm, they utilize first-order Taylor expansion and approximate the dual norm to obtain the ascent direction vector $\hat{\epsilon}$, optimizing the ascent direction loss to approximate Equation 2. As a result, the optimization problem of SAM reduces to the following equation:
\begin{equation}
\min _{\theta} \mathcal{L}(\theta+\hat{\epsilon} ; \mathcal{D}), \text { where } \hat{\epsilon} \triangleq \rho \frac{\nabla \mathcal{L}(\theta ; \mathcal{D})}{\|\nabla \mathcal{L}(\theta ; \mathcal{D})\|}
\end{equation}
For simplicity in notation, the loss function of SAM can be represented as follow:
\begin{equation}
	\mathcal{L}_{SAM}(\theta ; \mathcal{D})= \mathcal{L}(\theta+\hat{\epsilon} ; \mathcal{D})
\end{equation}

In the co-training scenario, SAM faces a challenge due to dataset size imbalances and the domain-specific artifacts. This can result in "learning shortcuts," where the model prioritizes learning from larger domains. Specifically, within each mini-batch in training, samples from the larger domain are more likely to be selected, which can result in some mini-batches not containing samples from the smaller domains. As a consequence, the ascent direction vector $\hat{\epsilon}$ may become skewed towards a larger domain ${L}$, which we term as domain ascent bias problem. To address this issue, we proposed co-training SAM (CSAM) method. 
CSAM define a specific data sampler according to the proportions of the datasets to ensure that each domain participated in the computation of the SAM loss function within every mini-batch. Assuming we have datasets from two domains, one with a larger number of data denoted as $L$, and the other with a smaller volume of data denoted as $S$. Then we will have the number of samples ${N}_{L}$ of domain $L$ and ${N}_{S}$ of domain $S$ in each mini-batch according to their proportion:
\begin{equation}
\begin{split}
{N}_{L} = \left\lfloor \frac{{len}_{L}}{{len}_{L} + {len}_{S}}  * {N}_{batch} \right\rfloor, \\
{N}_{S} = \left\lfloor \frac{{len}_{S}}{{len}_{L} + {len}_{S}}  * {N}_{batch} \right\rfloor,
\end{split}
\end{equation}
where ${{len}_{(\cdot)}}$ denotes the number of the domain dataset, $\left\lfloor\cdot	\right\rfloor$ represents a round down operation. 
The CSAM loss computation in each mini-batch can be calculate as follow: 
\begin{equation}
	\mathcal{L}_{CSAM}(\theta ; \mathcal{D}) =\sum_{l=1}^{{N}_{L}} \sum_{s=1}^{{N}_{S}} \ell \left(f\left(x_{l}^{L} + x_{s}^{S}  ; \theta + \hat{\epsilon}\right), y_{l}^{L} + y_{s}^{S}\right).
	\label{equation:loss}
\end{equation}
In each mini-batch, we ensured that each domain randomly appears according to its dataset proportion, ensuring that the ascent direction vector computes gradients from each domain in every mini-batch calculation, thus tackle the domain ascent bias issue.

\section{Experiments}
\subsection{Experiments setting}

We comprehensively conduct experiments on ADD baseline models under three conditions: Vocoder-trained (trained using vocoder-based audio), Codec-trained (trained using codec-based audio), and Co-trained (using both vocoder-based and codec-based audio).

At first, to validate whether training the ADD model using conventional vocoder-based audio can effectively detect codec-based audio, we construct the vocoder-trained ADD model using the training set of ASVspoof2019LA (19LA). 19LA was created utilizing 107 speakers (46 male, 61 female) utterances and contains a total number of 12,456 real and 108,978 fake speeches by using 11 TTS and 8 VC (A01-A19) spoofing algorithms. The training and development sets consist of data from spoofing systems denoted by A01 to A06, while the evaluation set includes attacks from spoofing systems denoted by A7 to A19. The details of 19LA can be found in Table \ref{tab:19LA}.

Then, we train the ADD model using the Codecfake training set 740,747 utterances and evaluate it on the testing conditions C1-C7 of Codecfake. 
While we observed significant improvements in performance compared to the vocoder-trained model under test conditions C1-C7, we also found that the codec-trained ADD model exhibited a decrease in performance when detecting vocoder-based audio. 
\begin{table}[!t]
	\caption{Summary of the 19LA dataset.}
	\label{tab:19LA}
	\centering
	\setlength{\tabcolsep}{1.5mm}
	\renewcommand{\arraystretch}{1.5}
	\begin{tabular}{|c|c|c|c|c|}
		\hline
		Subset &Fake Types &Real &Fake &Total\\
		\hline
		Training set &A01-A06 &2580 &22800 &25380\\
		\hline
		Development set  &A01-A06 &2548 &22296 &24844\\
		\hline
		Evaluation set &A07-A19 &7355 &63882 &71237\\
		\hline
	\end{tabular}
\end{table}
Thus, we attempted to train a universal ADD model using the co-trained dataset, aiming to detect both vocoder-based audio and codec-based audio. For the co-trained dataset, the training set comprises 766,127 samples, including 25,380 from the 19LA training set and 740,747 from the Codecfake training set. The validation set consists of 117,440 samples, with 24,844 from the 19LA development set and 92,596 from the Codecfake development set.

\begin{table*}[t]
	\caption{EER (\%) results for ADD methods trained by 19LA training set. CAVG represents the average EER across C1-C7.}
	\label{tab:vocoder-trained}
	\centering
	\setlength{\tabcolsep}{2mm} 
	\renewcommand{\arraystretch}{1.2}
	\begin{tabular}{|c|c|c|c|c|c|c|c|c|c|c|c|}
		\hline
		Model &19LA &ITW &C1&C2&C3&C4 &C5 &C6 &C7 &CAVG &AVG\\
		\hline
		Mel-LCNN  &5.084	&47.021	&36.127	&49.304	&\bf 39.234	&49.160	&46.991	&\bf 36.967	&49.584	&43.910 &39.941 \\
		\hline
		W2V2-LCNN  &0.625	&41.385	&\bf 19.481	&45.260	&45.010	&\bf 20.683	&\bf 34.321	&44.390	&\bf 43.717	& \bf 36.123 &\bf 32.764
		\\
		\hline
		WavLM-AASIST   &0.462	&\bf17.917	&41.752	&53.122	&50.469	&39.205	&44.224	&48.397	&47.838	&46.430
		&38.154
		\\
		\hline		
		W2V2-AASIST  &\bf 0.122	& 23.713	&40.142	&\bf 42.908	&44.564	&33.580	&39.197	&44.889	&45.804	& 41.583&34.991
		\\
		\hline
	\end{tabular}
	
\end{table*}

\begin{table*}[t]
	\caption{EER (\%) results for ADD methods trained by Codecfake training set. CAVG represents the average EER across C1-C7.}
	\label{tab:codec-trained}
	\centering
	\setlength{\tabcolsep}{2mm} 
	\renewcommand{\arraystretch}{1.2}
	\begin{tabular}{|c|c|c|c|c|c|c|c|c|c|c|c|}
		\hline
		Model &19LA &ITW &C1&C2&C3&C4 &C5 &C6 &C7 &CAVG &AVG\\
		\hline
		Mel-LCNN  &26.826	&42.635	&2.147	&9.563	&8.618	&12.239	&20.033	&8.951	&30.867	&13.203 &17.987 
		\\
		\hline
		W2V2-LCNN  &4.433	&\bf4.975	&\bf 0.030	&0.151	&0.144	&0.113	&0.635	&2.185	&5.178	&1.205 &1.983
		\\
		\hline
		WavLM-AASIST  &14.455	&22.528	&0.045	&0.121&0.839	&0.098	&0.076	&3.818	&15.270	&2.895 &6.361
		\\
		\hline		
		
		W2V2-AASIST  &\bf 3.806	&9.606	&0.167	&\bf 0.008	&\bf 0.023	&\bf 0.015	&\bf 0.038	&\bf 0.106	&\bf 0.884	&\bf 0.177 &\bf 1.628 
		\\
		\hline
	\end{tabular}
\end{table*}

\subsection{Implementation details}
For pre-processing of the ADD baseline models, all audio samples were first down-sampled to 16,000 Hz and trimmed or padded to a duration of 4 seconds. For the mel-spectrogram, we extracted a 80-dimensional mel-spectrogram. For self-supervised pre-trained feature, we employed the Wav2Vec-XLS-R\footnote{https://huggingface.co/facebook/wav2vec2-xls-r-300m} and WavLM-large\footnote{https://huggingface.co/microsoft/wavlm-large} with frozen parameters, extracting the 1024-dimensional hidden states as the feature representation. 

All models used Adam optimizer with a learning rate of 5$\times10^{-4}$. We conducted training for 100 epochs using weighted cross-entropy, with the weight for the real class set to 10 and for the fake class set to 1. The learning rate is halved every 10 epochs. For the codec-trained ADD model, training was performed for 10 epochs and the learning rate is halved every 2 epochs. The model that exhibited the best performance on the validation set was selected as the evaluation model. 

For experimental evaluation, we utilized the official implementation for EER calculation\footnote{https://github.com/asvspoof-challenge/2021/blob/main/eval-package/eval\_metrics.py}, maintaining precision up to three decimal places. For the computation of the confusion matrix, we utilized a threshold of 0.5 for discriminating between real and fake predictions and conducted the calculations from SKlearn.

\section{Results and Disscussion}
\subsection{Vocoder-trained ADD model}
First, we explore whether the vocoder-trained ADD model can effectively detect codec-based audio, as outlined in Table \ref{tab:vocoder-trained}. Concretely, we conducted training on four baseline models: Mel-LCNN, W2V2-LCNN, WavLM-AASIST and W2V2-AASIST, utilizing the ASVspoof2019LA (19LA) \cite{todisco19_interspeech} training set. From the results, we can first observe that they achieve promising result in 19LA test sets and W2V2-AASIST achieve the lowest EER with 0.122\%. As the 19LA training set incorporates six spoofing methods, and the test set encompasses a total of unseen 13 methods, achieving favorable results on the 19LA test set suggests that the detection system exhibits generalizability. However, when the vocoder-trained ADD models evaluate on the Codecfake test sets, the results are quite unfavorable. Mel-LCNN, W2V2-LCNN, WavLM-AASIST and W2V2-AASIST achieve average EER of 39.941\%, 32.764\%, 38.154\% and 34.991\%, respectively, which represents a decrease in performance compared to the 19LA test set. This indicates that models trained with vocoders cannot generalize to effectively detect codec-based audio. 

\begin{figure*}[!t]
	\centering
	\subfloat{\includegraphics[width=6in]{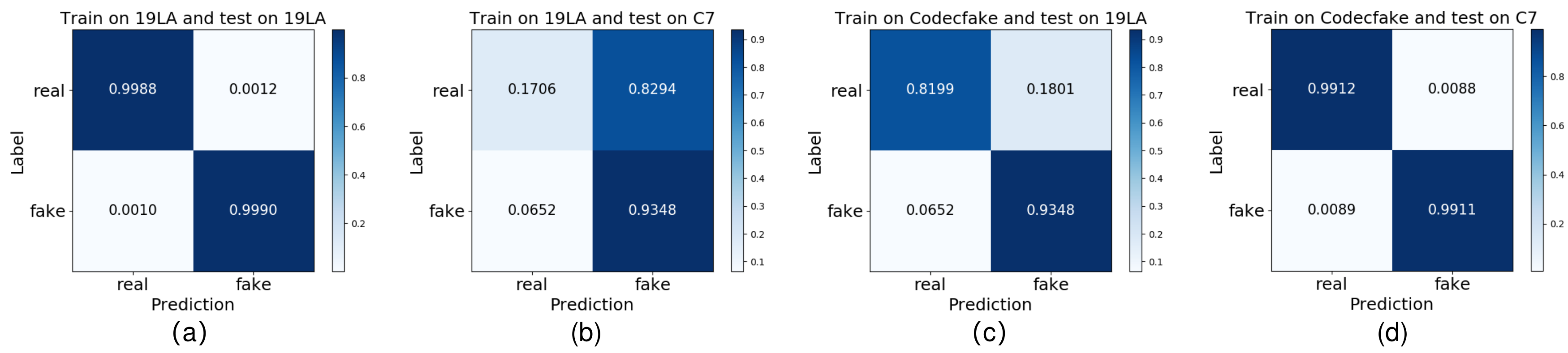}}
	\hfil
	\caption{The confusion matrices under different test conditions. (a), (b) correspond to W2V2-AASIST trained on the 19LA training set and tested on 19LA and C7. (c), (d) correspond to W2V2-AASIST trained on the Codecfake training set and tested on 19LA and C7.}
	\label{fig:confusion} 
\end{figure*}

\subsection{Codec-trained ADD model}
Due to the poor performance of vocoder-trained ADD models on the Codecfake test set, we proceed to train the baseline models using the Codecfake training set. The results are presented in the Table \ref{tab:codec-trained}. Mel-LCNN, W2V2-LCNN, WavLM-AASIST and W2V2-AASIST achieve average EER of 17.987\%, 1.983\%, 6.361\% and 1.628\%, exhibited reductions in average EER of 21.954\%, 30.781\%, 31.793\% and 33.363\%, respectively, compared to the vocoder-trained model. 
All ADD models trained with the Codecfake dataset exhibit promising results under Codecfake test condition C1-C7. Among these, W2V2-AASIST achieved the best average in EER of 0.177\%. In the unseen codec test condition C7, W2V2-AASIST achieves an EER of 0.884\%, indicating that training with the Codecfake dataset can generalize to detecting unknown codec-based audio. Furthermore, we found that using only the codec-trained ADD model can detect vocoder-based audio. Specifically, W2V2-AASIST achieves a 3.806\% EER on the 19LA test set, and W2V2-LCNN achieves a 4.975\% EER on the ITW dataset. For the ITW results, as far as we know, they already surpass all the ADD models trained on vocoder-based audio.

Through the experiments, we observed a decrease in performance of vocoder-trained ADD models when testing codec-based audio, and similarly, a decrease in performance of codec-trained ADD models when testing vocoder-based audio such as test on 19LA. To delve deeper into the factors contributing to the degraded performance of ADD models under these OOD test conditions, we plotted the confusion matrix graphs in Figure \ref*{fig:confusion}. Figure \ref*{fig:confusion}(a) illustrates the performance of vocoder-trained W2V2-AASIST on 19LA, where both real and fake samples are almost accurately predicted. However, Figure \ref*{fig:confusion}(b) shows its performance on C7, revealing a significant number of real audio samples being misclassified as fake, leading to a decrease in EER. In Figure \ref*{fig:confusion}(c) and Figure \ref*{fig:confusion}(d), the codec-trained ADD model demonstrates an enhanced ability to distinguish between real and fake audio. This is particularly evident with codec-based audio, which is inherently more authentic than vocoder-based audio, leading to superior performance in both the out-of-distribution scenario of 19LA and the unseen condition of C7.

\begin{table*}[t]
	\caption{EER (\%) results for ADD methods trained by Codecfake training set and test on different parameter setting on C3 and C4.}
	\label{tab:parameter}
	\centering
	\setlength{\tabcolsep}{2mm} 
	\renewcommand{\arraystretch}{1.2}
	\begin{tabular}{|c|c|c|c|c|c|c|c|c|c|c|c|}
		\hline
		Model &C3-1 &C3-2 &C3-3 &C3-4 &C4-1 &C4-2 &C4-3&AVG\\
		\hline
		Mel-LCNN  &8.180	&8.497	&8.618	&8.618	&13.252	&13.056	&16.836 &11.008
		\\
		\hline
		W2V2-LCNN  &0.083	&0.144	&0.144	&0.144	&0.038 &0.786	&2.192	&0.504
		\\
		\hline
		WavLM-AASIST &1.769	&2.480	&2.744	&2.782	&0.045	&0.386	&1.126	&1.619
		\\
		\hline
		W2V2-AASIST &\bf 0.015	&\bf 0.015	&\bf 0.015	&\bf 0.015	&\bf 0.008	&\bf 0.166	&\bf 0.650	&\bf 0.126
		\\
		\hline
	\end{tabular}
	
\end{table*}

\begin{table*}[t]
	\caption{EER (\%) results for the impact of language setting. All experiments use the top-performing ADD model, W2V2-AASIST. 'English' refers to training or testing with the VCTK subset of Codecfake, while 'Chinese' denotes the ALSHELL subset. 'ALL' indicates the original multilingual testing conditions of Codecfake.}
	\label{tab:language}
	\centering
	\setlength{\tabcolsep}{2mm} 
	\renewcommand{\arraystretch}{1.2}
	\begin{tabular}{|c|c|c|c|c|c|c|c|c|c|c|}
		\hline
		Train & Test & C1 & C2 & C3 & C4 & C5 & C6 & C7 & AVG \\
		\hline
		English & English & \bf 0.000 & \bf 0.000 & \bf 0.045 & \bf 0.022 & \bf 0.000 & \bf 0.247 & \bf 1.123 & \bf 0.205 \\
		\hline
		English & Chinese & 0.763 & 1.219 & 1.618 & 3.509 & 1.128 & 4.295 & 14.584 & 3.874 \\
		\hline
		English & ALL & 0.302 & 0.605 & 1.225 & 1.875 & 0.514 & 4.445 & 10.425 & 2.770 \\
		\hline
		Chinese & English & 0.135 & 0.764 & 3.595 & 0.247 & 0.247 & 8.627 & 14.626 & 4.034 \\
		\hline
		Chinese & Chinese & 0.182 & 0.057 & 0.330 & 0.091 & 0.046 & 1.686 & 15.096 & 2.498 \\
		\hline
		Chinese & ALL & 1.051 & 0.431 & 7.998 & 0.370 & 0.227 & 12.799 & 26.293 & 7.024 \\
		\hline
	\end{tabular}
\end{table*}
\begin{table*}[t]
	\caption{EER (\%) results for ADD methods test on ALM-based audio.}
	\label{tab:ALM}
	\centering
	\setlength{\tabcolsep}{2mm} 
	\renewcommand{\arraystretch}{1.2}
	\begin{tabular}{|c|c|c|c|c|c|c|c|c|c|c|c|}
		\hline
		\multirow{2}{*}{Model}&\multicolumn{4}{c|}{Vocoder-trained ADD model} & \multicolumn{4}{c|}{Codec-trained ADD model}\\
		
		\cline{2-9}
		&A1 &A2 &A3 &AVG &A1 &A2 &A3&AVG\\
		\hline
		Mel-LCNN  &21.233	&3.241	&\bf31.847	&18.774
		&7.393	&9.553	&\bf27.254	&14.733
		
		\\
		\hline
		W2V2-LCNN  &0.608	&2.093	&49.284	&17.328
		&0.450	&1.080	&37.076	&12.869
		\\
		\hline		
		WavLM-AASIST &0.495	&1.733	&66.207	&22.812
		&\bf 0.113	&3.364	&36.975	&13.484
		
		\\
		\hline
		W2V2-AASIST &\bf0.225	&\bf 0.833	&45.674	&\bf 15.577
		&0.225	&\bf 0.135	&33.803	&\bf 11.388
		
		\\
		\hline
	\end{tabular}
	
\end{table*}

\subsection{Impact of Codec Settings}
To further validate different codec parameters setting will have influence performance of ADD model, we tested various configurations of C3 and C4 using a codec-trained ADD model as shown in Table \ref{tab:parameter}. It can be observed that compared to the original settings of C3 and C4, using different parameters to generate codec-based audio does not significantly impact the performance of the anti-spoofing system. Especially for C3-FuncCodec, the EERs of C3-1 to C3-4 across the three ADD systems are almost identical. For C4-Encodec, it can be observed that from C4-1 to C4-3, the EER of the ADD model gradually increases. This indicates that using higher bps and quantizers in codec synthesis system tends to produce more authentic audio, making it more challenging for the ADD model to distinguish.

\subsection{Impact of language setting}
Codecfake is generated based on two source languages: Chinese from AISHELL and English from VCTK. In this section, we explore the impact of language on ADD models. We conducted the following experiments, first dividing the Codecfake training set, development set, and evaluation set based on AISHELL and VCTK. The experiments are categorized into intra-language testing and cross-language testing. In intra-language testing, both the training and testing languages are kept consistent. While in cross-language testing, the training and testing languages differ, with the experiment results presented in Table \ref{tab:language}. 

From the experiments results, we can draw several conclusions. First, cross-language testing results in a significant performance drop compared to intra-language testing. For example, the EER for English training with Chinese testing worsens by 3.669\% compared to English training with English testing, and the EER for Chinese training with English testing worsens by 1.536\% compared to Chinese training with Chinese testing. Second, the ADD model trained solely on English outperforms the model trained on Chinese. When testing on the complete set of Codecfake test conditions, the model trained in English achieves an EER of 2.770\%, while the model trained in Chinese achieves an EER of 7.024\%. Additionally, in intra-language testing, training and testing in English achieve an EER of 0.205\%, which is the best result in language testing experiments. Finally, the performance of multilingual training surpasses that of training solely on one language. By utilizing the entire multilingual Codecfake training set, the CAVG achieves an EER of 0.177\%, significantly lower than when using only a single language for training. This demonstrates the advantages of the multilingual design of Codecfake.

\begin{table*}[t]
	\caption{EER (\%) results for ADD methods co-trained by 19LA training set and Codecfake training set.}
	\label{tab:eer-cotrained}
	\centering
	\setlength{\tabcolsep}{2mm} 
	\renewcommand{\arraystretch}{1.2}
	\begin{tabular}{|c|c|c|c|c|c|c|c|c|c|c|c|}
		\hline
		Model &19LA &ITW &C1&C2&C3&C4 &C5 &C6 &C7 &CAVG &AVG\\
		\hline
		Mel-LCNN  &10.214	&36.757	&6.675	&22.052	&7.310	&10.410	&15.906	&7.620	&30.307	&14.326
		&16.361
		
		\\
		\hline
		W2V2-LCNN  &1.008	&5.847	&0.295	&2.480	&2.540	&4.203	&19.965 &23.949	&34.563	&12.571
		&10.539
		
		\\
		\hline
		WavLM-AASIST &1.072	&23.264	&0.083	&0.045	&0.355	&0.038	&0.015	&1.451	&11.779	&1.967
		&4.234
		
		\\
		\hline		
		W2V2-AASIST  &0.625	&7.184	&0.015	&0.030	&0.023	&0.023	&0.038	&0.098&	0.627	&0.122 &0.963\\
		\hline
		W2V2-AASIST(SAM)  &\bf 0.231	&7.483	&\bf0.008	&\bf0.008	&\bf0.000	&\bf0.000	&\bf0.015	&0.038	&0.696	&0.109 &0.942\\
		\hline
		W2V2-AASIST(ASAM)  &0.314	&9.994	&0.015	&0.030	&0.015	&0.008	&0.023	&0.060	&0.711	&0.123 &1.241\\
		\hline
		W2V2-AASIST(CSAM)  &0.313	&\bf4.689	&\bf0.008	&0.023	&0.008	&0.008	&0.030	&\bf0.030	&\bf0.431	&\bf0.077 &\bf 0.616\\
		\hline
	\end{tabular}
	
\end{table*}

\subsection{Test on ALM-based audio}
In practice, ALM-based audio encompasses not only the reconstruction by neural codec method but also manipulation in content. To assess whether ADD models trained on the codecfake dataset can effectively detect ALM-based audio, we conducted experiments as shown in the Table \ref{tab:ALM}. At first, we observed that employing solely vocoder-trained ADD models yielded satisfactory outcomes for ALM-based audio, particularly evident in evaluations on A1 and A2, where models such as W2V2-AASIST achieved EERs below 1\%. However, substantial performance deterioration was evident in evaluations on the condition A3, attributable to OOD issues. The ALM test condition A3 is characterized by sounds with event descriptions that deviate from the distribution of speech. After using Codecfake to train ADD model, performance across A1-A3 are improved, with W2V2-AASIST achieving the lowest average EER of 11.388\%, representing a decrease of 4.189\% compared to the vocoder-trained ADD models. However, it is noteworthy that while the EER for A3 decreased compared to the vocoder-trained ADD models, it still remains relatively high.
\begin{figure}[!t]
	\centering
	\subfloat{\includegraphics[width=2.5in]{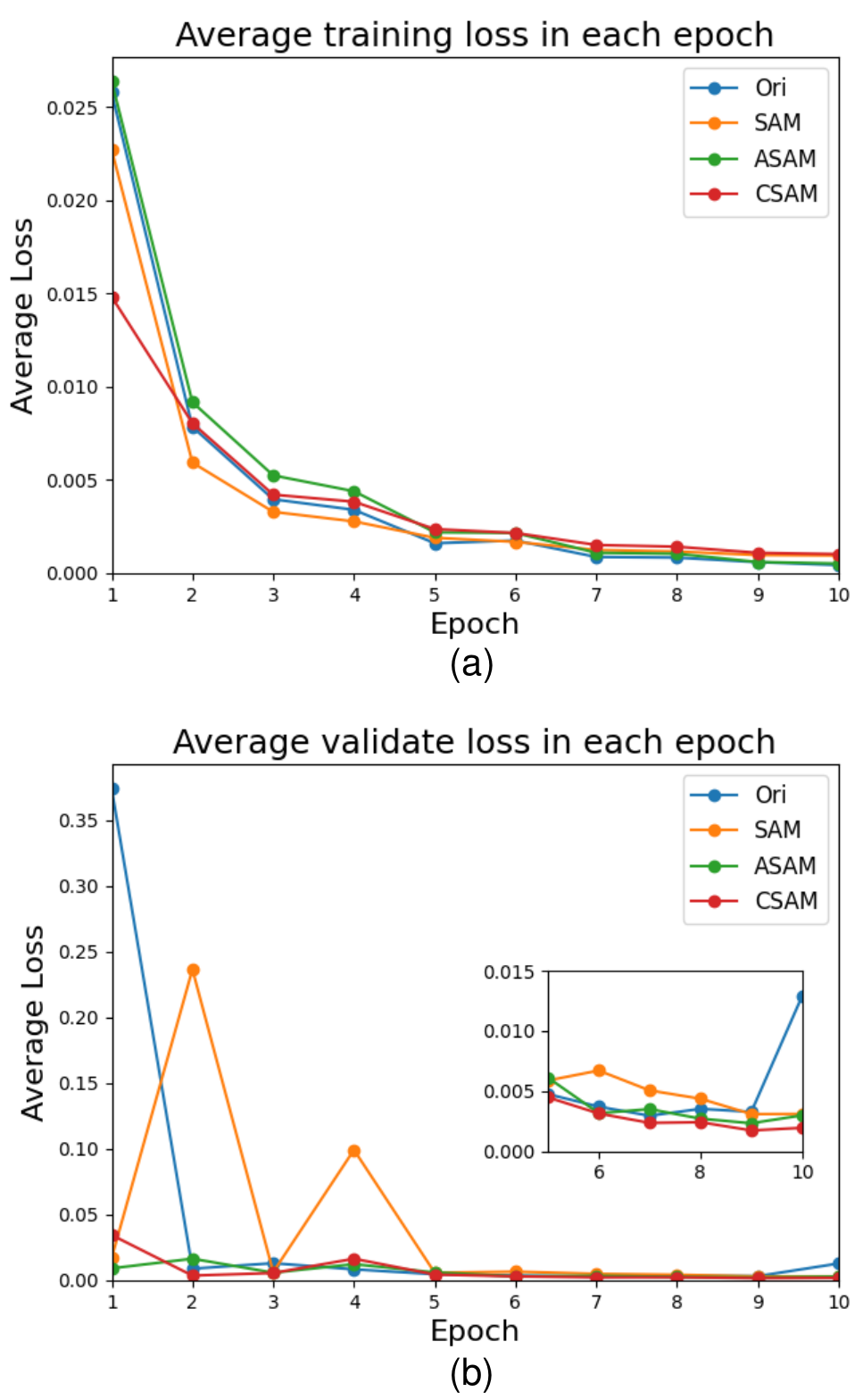}}
	\hfil
	\caption{Average loss in each epoch during the training and validation processes. (a) and (b) correspond to the co-trained W2V2-AASIST during training and validation, respectively.}
	\label{fig:loss} 
\end{figure}
\subsection{Countermeasure for generalized ADD method}
Through the experiments described above, we observed a phenomenon: while training in different domains yields optimal performance in their respective test domains, it leads to performance degradation in cross-domain test scenarios. In this section, we propose a co-training approach using vocoder-based audio and codec-based audio to ensure that the detection model performs well across all scenarios. Furthermore, during the co-training process, we employed the proposed CSAM method. This approach optimizes the ERM while learning flat minima, avoiding the domain ascent bias problem that may arise during the co-train situation in SAM. Experiment result is shown in Table \ref{tab:eer-cotrained}, where SAM is original implementation, ASAM\cite{kwon2021asam} is a modified SAM approach to improve generalization performance by parameter re-scaling, CSAM is our proposed approach. 
Firstly, the performance of the co-trained ADD model has improved compared to both vocoder-trained ADD model and codec-trained ADD model under the testing conditions. For instance, the top-performing model, W2V2-AASIST, achieved a CAVG of 0.122\% and an AVG EER of 0.963\%, compared to the vocoder-trained W2V2-AASIST with a CAVG of 41.583\% and an AVG EER of 34.991\%, and the codec-trained W2V2-AASIST with a CAVG of 0.177\% and an AVG EER of 1.628\%, indicating a decrease in EER. Then, to further enhance the model's generalizability, we trained all model in the co-training scenario with SAM, ASAM, and CSAM strategy. 
After employing the original SAM, it was observed that apart from an increase in EER on the ITW dataset and the C7 unseen test set, there was a decrease in EER across all other test scenarios. Particularly noteworthy were the remarkable scores of 0\% EER obtained under the C3 and C4 test conditions. After further employing CSAM, compared to original SAM, although there was a slight increase in EER on the visible test condition, there was a further improvement in the performance on unseen OOD test conditions. This is because SAM tends to overfit on a specific ID domain during multi-domain training, whereas CSAM computes each mini-batch across all domains, which can significantly enhances OOD performance. Specifically, using W2V2-AASIST (CSAM) resulted in a decrease in EER from 7.483\% to 4.689\% on the ITW test condition and from 0.627\% to 0.431\% on the unseen C7 test set, compared to W2V2-AASIST (SAM). Finally, W2V2-AASIST (CSAM) achieved the lowest average EER, with a CAVG of 0.077\% and an AVG of 0.616\%.

To further investigate whether SAM benefits the model training, we plotted the loss curves of co-trained W2V2-AASIST during training and validation as shown in the Figure \ref{fig:loss}. From Figure \ref{fig:loss}(a), 
we observed that the convergence process of loss is similar across different SAM methods. However, CSAM exhibits lower loss values as early as the first epoch, indicating that the CSAM approach can accelerate the convergence speed of ERM. 
From Figure \ref{fig:loss}(b), in the validation process, whether SAM, ASAM, or CSAM is applied, there is a noticeable decrease in loss compared to when SAM is not used, especially in the first epoch. Furthermore, for epochs 5-10, we observe that without SAM, the seventh epoch is optimal for the model, while the SAM series consistently optimizes the validation loss, achieving the best model in the ninth epoch. When training a ADD model, even if the 19LA task runs for 100 epochs, the optimal checkpoint often emerges around epochs 20-30. Subsequent reductions in the learning rate are insufficient to learn a more generalized model on the validation set. However, SAM optimizes ERM while simultaneously learning flat minima, allowing each epoch to continuously seek the most generalizable solution. In the validation set, CSAM achieves the lowest loss in the ninth epoch, corresponding to its best performance on the test set.

\section{Limitation and Future direction}
\label{section:Limitation}
Through our experiments, we have identified some limitations and areas for further improvement in our work. Future efforts will focus on the following items to make enhancements.

{{\bf More diverse audio type.} Codecfake dataset now incorporates mainstream codec methods and few of ALM-based method in speech. However, the performance of ADD models significantly declines when dealing with non-speech tasks, such as in the A3 scenario. To further develop a universally applicable ADD model, the dataset should not be limited to speech alone. Current methods like ALM can also synthesize other audio types, such as singing voices and audio events. Therefore, ADD dataset needs to include these different types of audio to achieve universally detection. }
	
{{\bf More acoustic condition.} The current synthesized audio in the Codecfake dataset only includes clean conditions, which may lack robustness for detection in complex real-world noise environments. To address this limitation and training a robust ADD model for real-world scenarios, it is necessary to apply various reverberation and noise enhancements to Codecfake dataset.}
	
{{\bf Generalized ADD methods.} The current ADD methods, such as W2V2-AASIST, exhibit a significant performance decline in OOD scenarios. In this paper, we attempted to achieve a generalized ADD model using co-training and CSAM approaches. However, there remains significant for improvement in OOD scenarios such as ITW and A3. Designing a generalized approach that can detect both vocoder-based audio and ALM-based audio is currently a crucial challenge.}	
	
{{\bf Source tracing methods.} It is crucial to develop source tracing methods, as this significantly contributes to safeguarding the copyright of ALM developers. Specifically, we need to design a source tracing method with generalization capabilities. This method should be able to perform classification in the ID ALM and identify novel ALM in OOD situations, which can effectively addressing the continuous emergence of ALM algorithms.}
		
\section{Conclusions}
Unlike existing vocoder-based deepfake audio datasets, in this paper, we propose the Codecfake dataset for effectively detecting novel ALM-based audio. Codecfake contains seven representative codec synthesis methods, including two languages, millions of audio samples, and various test conditions. Through experiments, we first verified that vocoder-trained ADD models are insufficient to detect the currently novel codec-based audio. Then, ADD models trained on the Codecfake dataset can correctly detect codec-based audio. To effectively detect both vocoder-based audio and codec-based audio, we propose the CSAM training strategy which can optimize both ERM while learning flat minima. Our work for detecting ALM-based audio is preliminary, and some limitations still exist. Future endeavors will aim to address these aforementioned limitations.

\bibliographystyle{IEEEbib}

\bibliography{myrefs} 
\vspace{2pt}

\begin{IEEEbiography}
	[{\includegraphics[width=1in,height=1.25in,clip,keepaspectratio]{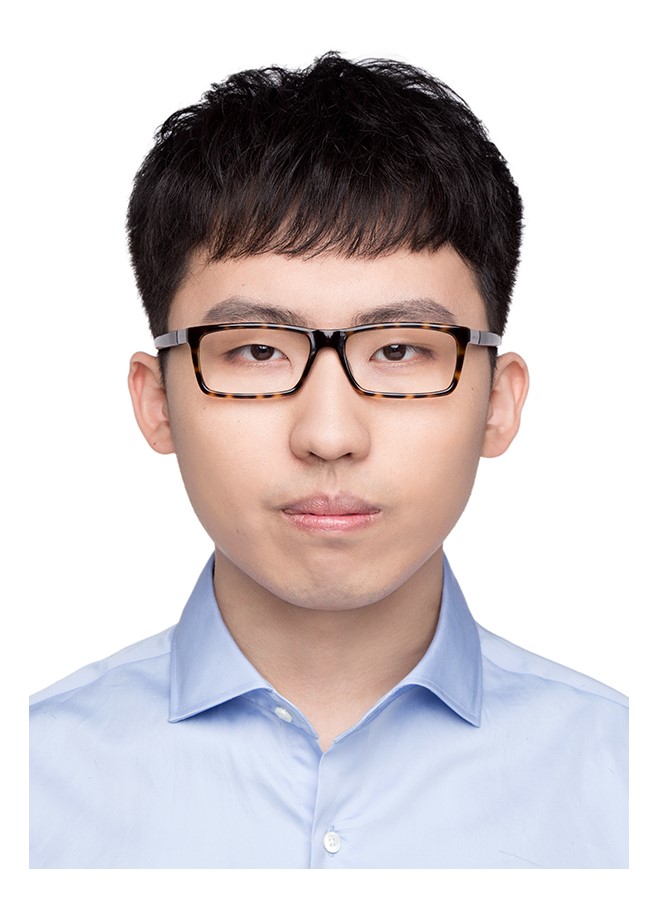}}]{Yuankun Xie}
	received the M.S. degree in communication and information system from Communication University of China, Beijing, in 2021, where he is currently pursuing the Ph.D. degree with the School of Information and Communication Engineering, Communication University of China, Beijing. He is also a joint Ph.D. student with the Institute of Automation, Chinese Academy of Sciences, Beijing. His research interests include audio deepfake detection, domain generalization, out-of-distribution detection, audio neural codec. 
\end{IEEEbiography}

\begin{IEEEbiography}
	[{\includegraphics[width=1in,height=1.25in,clip,keepaspectratio]{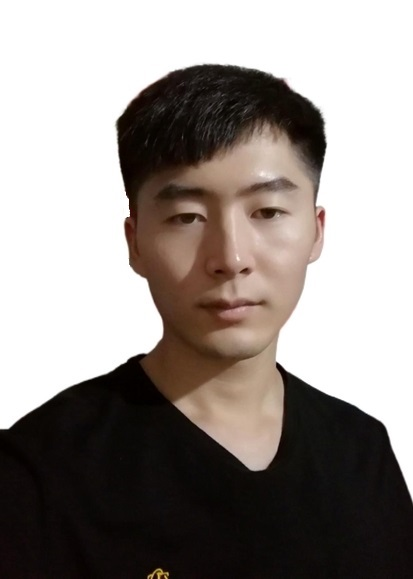}}]{Yi Lu}
	graduated from Shandong University of Science and Technology in 2021 with a bachelor's degree in Software Engineering from the School of Computer Science and Engineering. He is currently pursuing a master's degree at the School of Artificial Intelligence, University of Chinese Academy of Sciences. His research interests focus on Text-to-Speech, with a particular emphasis on language model-based speech synthesis and personalized speech synthesis.
\end{IEEEbiography}

\begin{IEEEbiography}
	[{\includegraphics[width=1in,height=1.25in,clip,keepaspectratio]{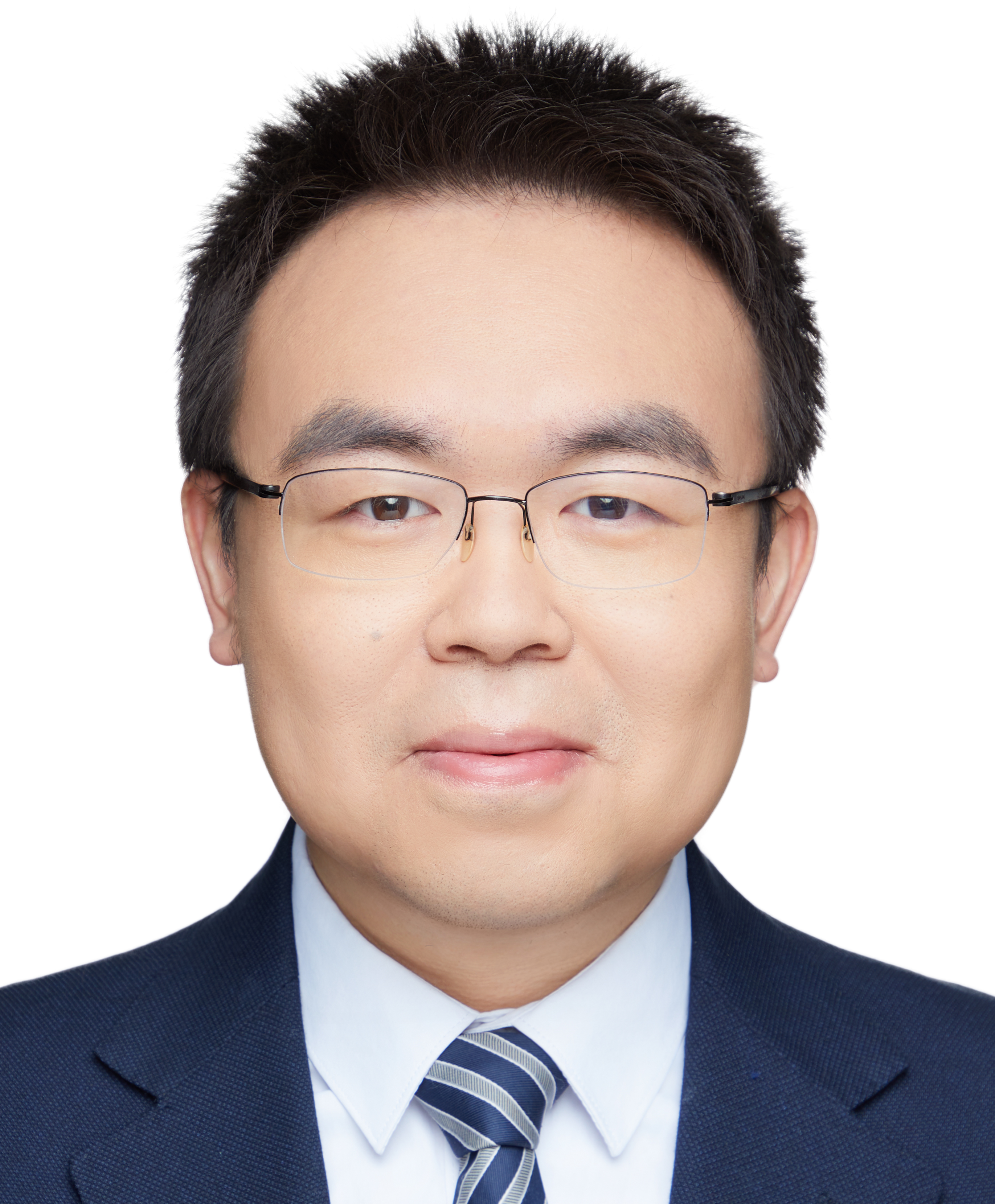}}]{Ruibo Fu}
	is an Associate professor in the National Laboratory of Pattern Recognition, Institute of Automation, Chinese Academy of Sciences, Beijing. He obtained B.E. from Beijing University of Aeronautics and Astronautics in 2015 and Ph.D. from the Institute of Automation, Chinese Academy of Sciences in 2020. His research interest is speech synthesis and transfer learning. He has published more than 20 papers in international conferences and journals such as ICASSP and INTERSPEECH and has won the best paper award twice in NCMMSC 2017 and 2019. He won the first prize in the personalized speech synthesis competition held by the Ministry of Industry and Information Technology twice in 2019 and 2020. He also won the first prize in the ICASSP2021 Multi-Speaker Multi-Style Voice Cloning Challenge (M2VoC) Challenge.
\end{IEEEbiography}

\begin{IEEEbiography}
	[{\includegraphics[width=1in,height=1.25in,clip,keepaspectratio]{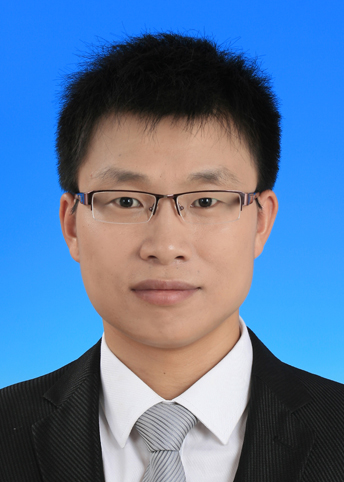}}]{Zhengqi Wen}
	(Member, IEEE) received the B.S. degree from the University of Science and Technology of China, Hefei, China, in 2008, and the Ph.D. degree from the Chinese Academy of Sciences, Beijing, China, in 2013, both in pattern recognition and intelligent system. He is currently an Associate Professor with the Beijing National Research Center for Information Science and Technology, Tsinghua University, Beijing, China. His current research interests include speech processing, speech recognition, and speech synthesis.
	
\end{IEEEbiography}

\begin{IEEEbiography}
	[{\includegraphics[width=1in,height=1.25in,clip,keepaspectratio]{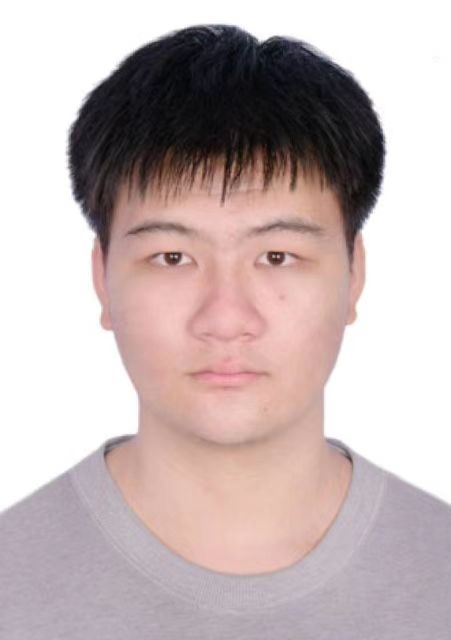}}]{Zhiyong Wang}
	received the B.E. degree in information management and information system from the Changsha University of Science and Technology, Changsha, China, in 2021. He is currently working toward the M.S. degree with the School of Artificial Intelligence, University of Chinese Academy of Sciences. His research interests include audio deepfake detection and speech to speech generation.
\end{IEEEbiography}

\begin{IEEEbiography}
	[{\includegraphics[width=1in,height=1.25in,clip,keepaspectratio]{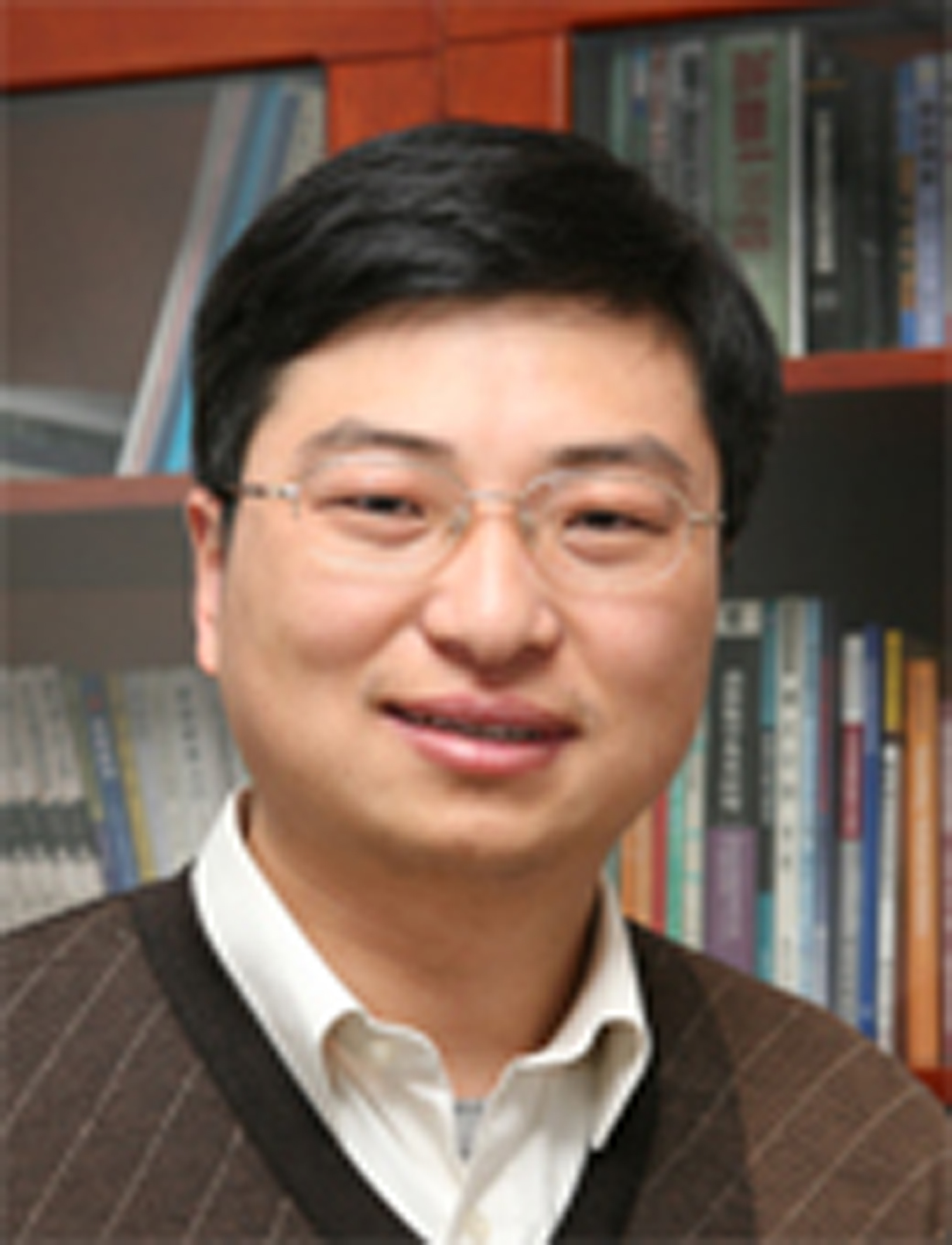}}]{Jianhua Tao}
	(Senior Member, IEEE) received the M.S. degree from Nanjing University, Nanjing, China, in 1996, and the Ph.D. degree from Tsinghua University, Beijing, China, in 2001. He is currently a Professor with Department of Automation, Tsinghua University, Beijing, China. He has authored or coauthored more than 300 papers on major journals and proceedings including the IEEE TASLP, IEEE TAFFC, IEEE TIP, IEEE TSMCB, Information Fusion, etc. His current research interests include speech recognition and synthesis, affective computing, and pattern recognition. He is the Board Member of ISCA, the chairperson of ISCA SIG-CSLP, the Chair or Program Committee Member for several major conferences, including Interspeech, ICPR, ACII, ICMI, ISCSLP, etc. He was the subject editor for the Speech Communication, and is an Associate Editor for Journal on Multimodal User Interface and International Journal on Synthetic Emotions. He was the recipient of several awards from important conferences, including Interspeech, NCMMSC, etc.
\end{IEEEbiography}

\begin{IEEEbiography}
	[{\includegraphics[width=1in,height=1.25in,clip,keepaspectratio]{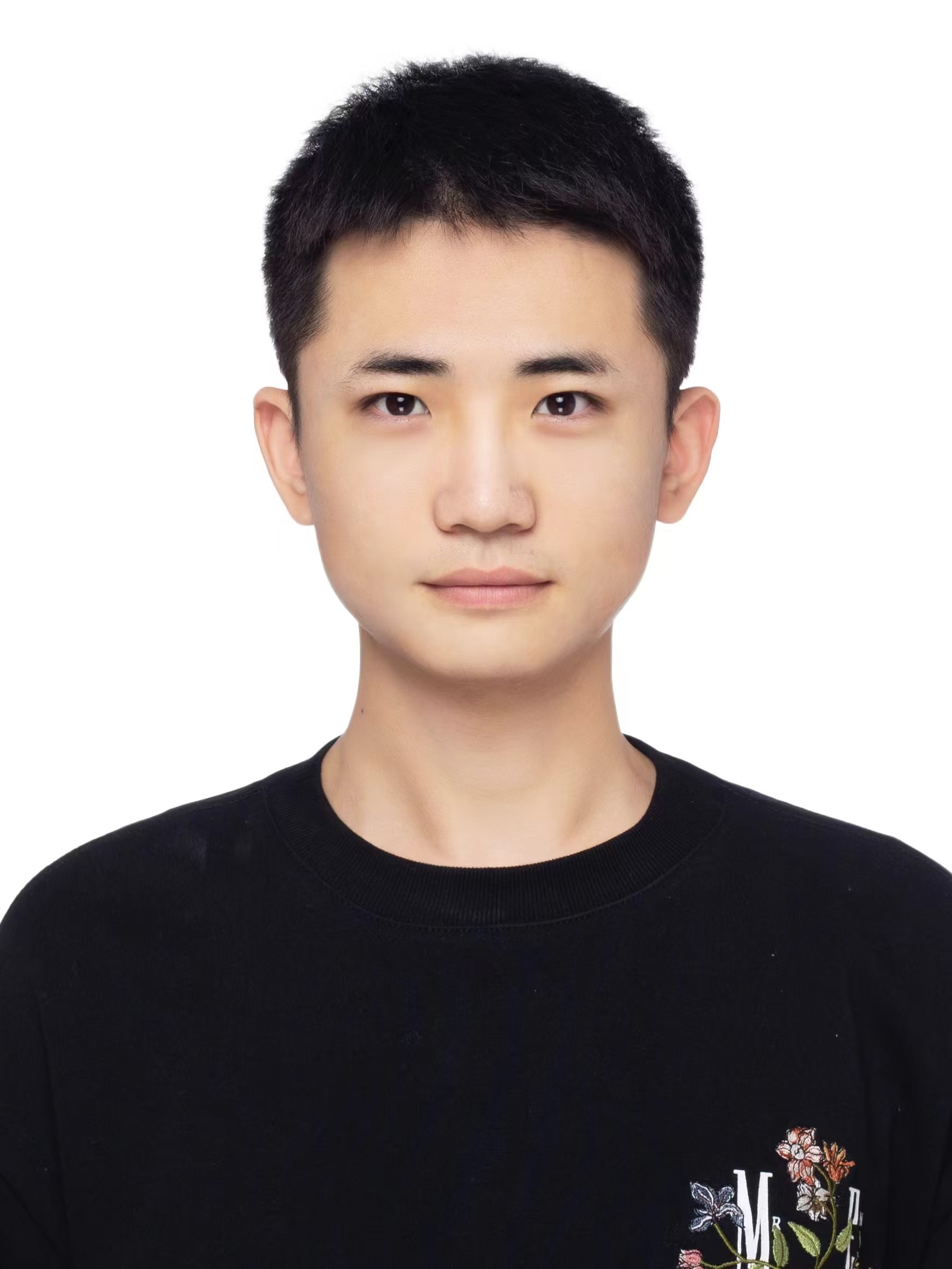}}]{Xiaopeng Wang}
	graduated from Southwest Minzu University in 2022 with a bachelor's degree in Computer Science and Technology. He is currently pursuing the master's degree at the School of Artificial Intelligence, University of Chinese Academy of Sciences. His main research areas are deepfake detection and text-to-speech.
\end{IEEEbiography}

\begin{IEEEbiography}
	[{\includegraphics[width=1in,height=1.25in,clip,keepaspectratio]{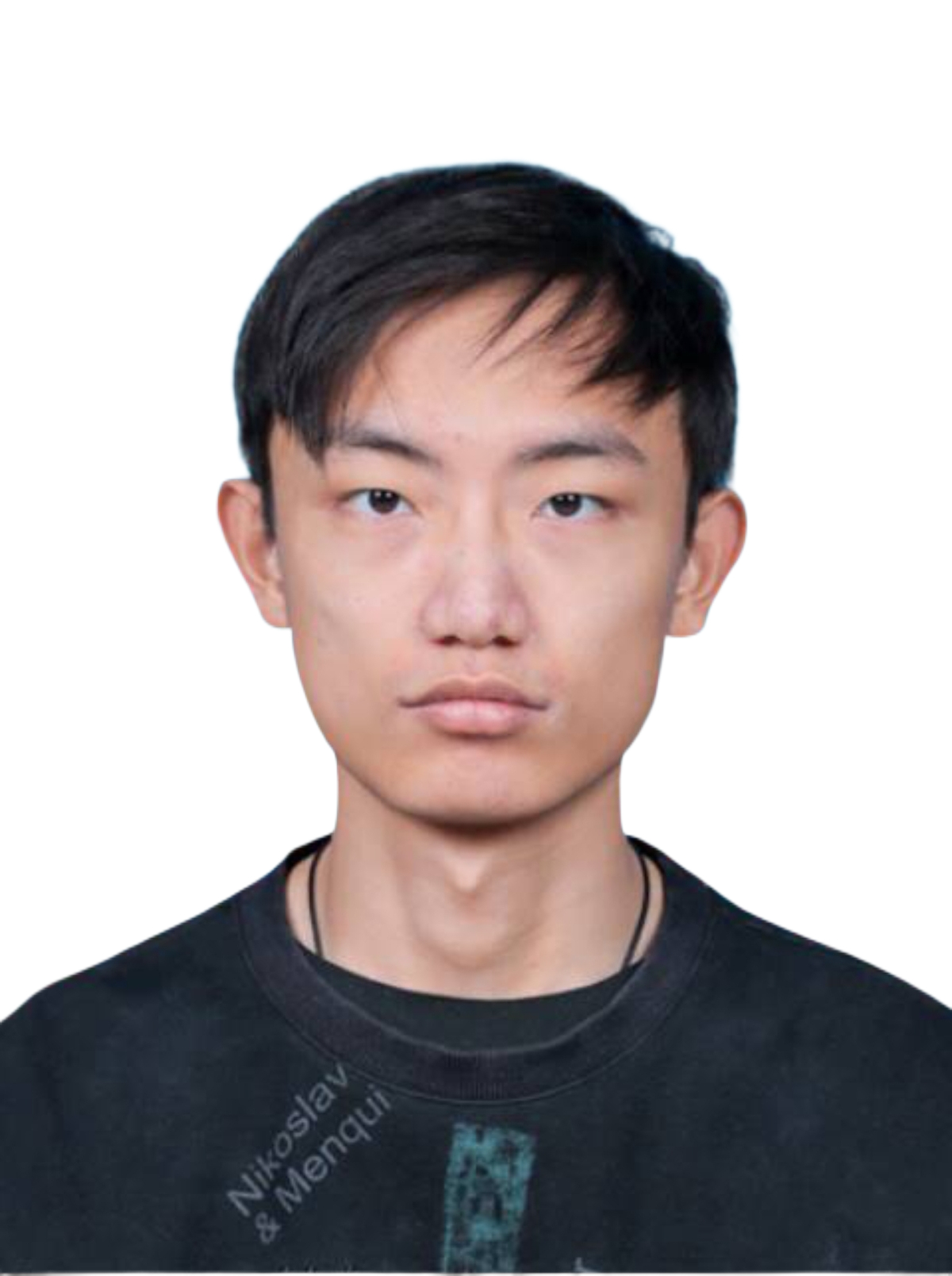}}]{Xin Qi}
	graduated from Shandong University of Science and Technology in 2021 with a bachelor's degree in Software Engineering. He is currently pursuing a master's degree at the School of Artificial Intelligence, University of Chinese Academy of Sciences, with a main research focus on text-to-speech technology.
\end{IEEEbiography}

\begin{IEEEbiography}
	[{\includegraphics[width=1in,height=1.25in,clip,keepaspectratio]{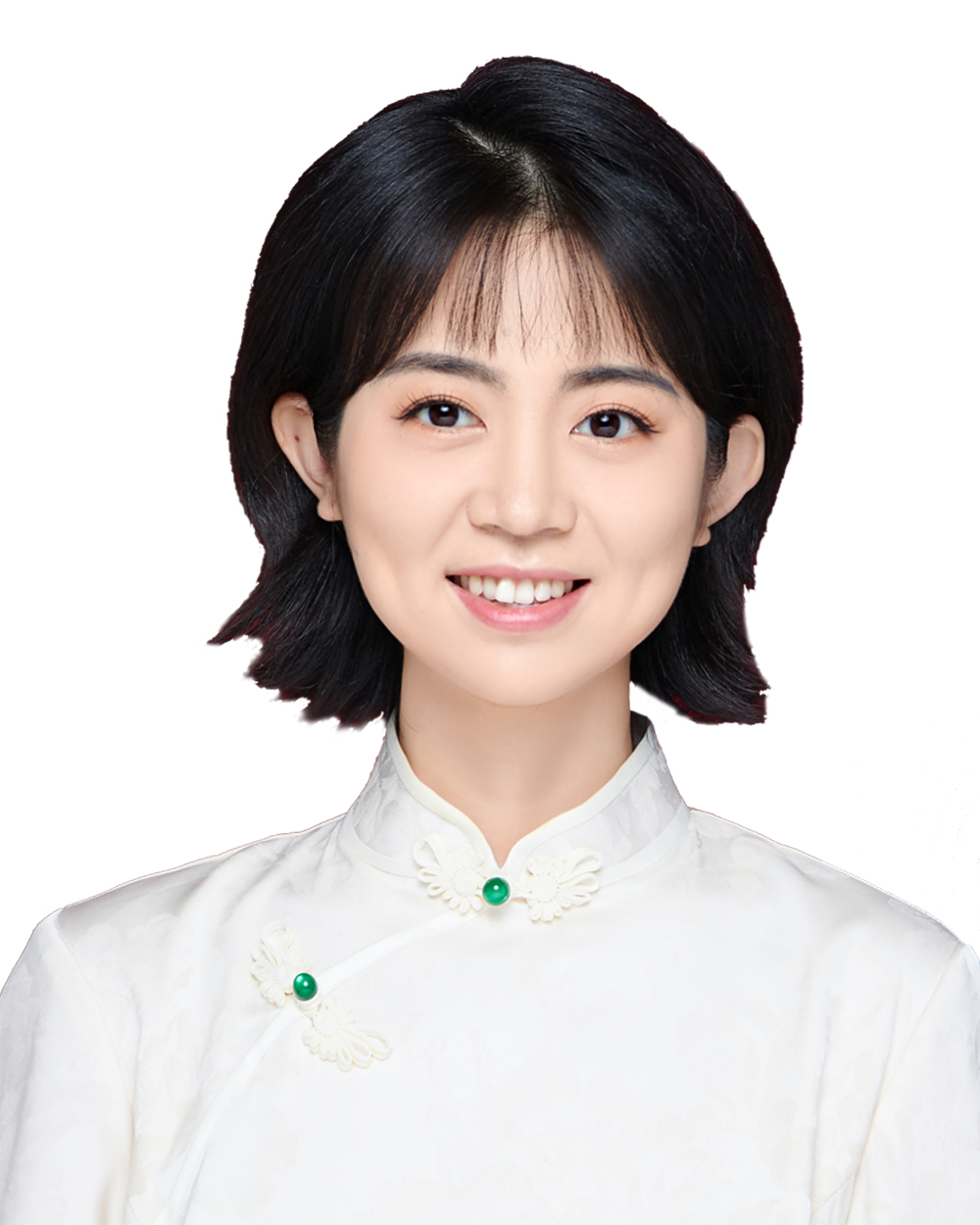}}]{Haonan Cheng}
	received the bachelor's and Ph.D. degrees in computer science from Tianjin University, Tianjin, China, in 2016 and 2021, respectively. She is currently an Associate Professor with the State Key Laboratory of Media Convergence and Communication, Communication University of China, Beijing, China. Her research interests include cross-modal sound generation, audio processing, music information retrieval, computer graphics and virtual reality.
\end{IEEEbiography}

\begin{IEEEbiography}
	[{\includegraphics[width=1in,height=1.25in,clip,keepaspectratio]{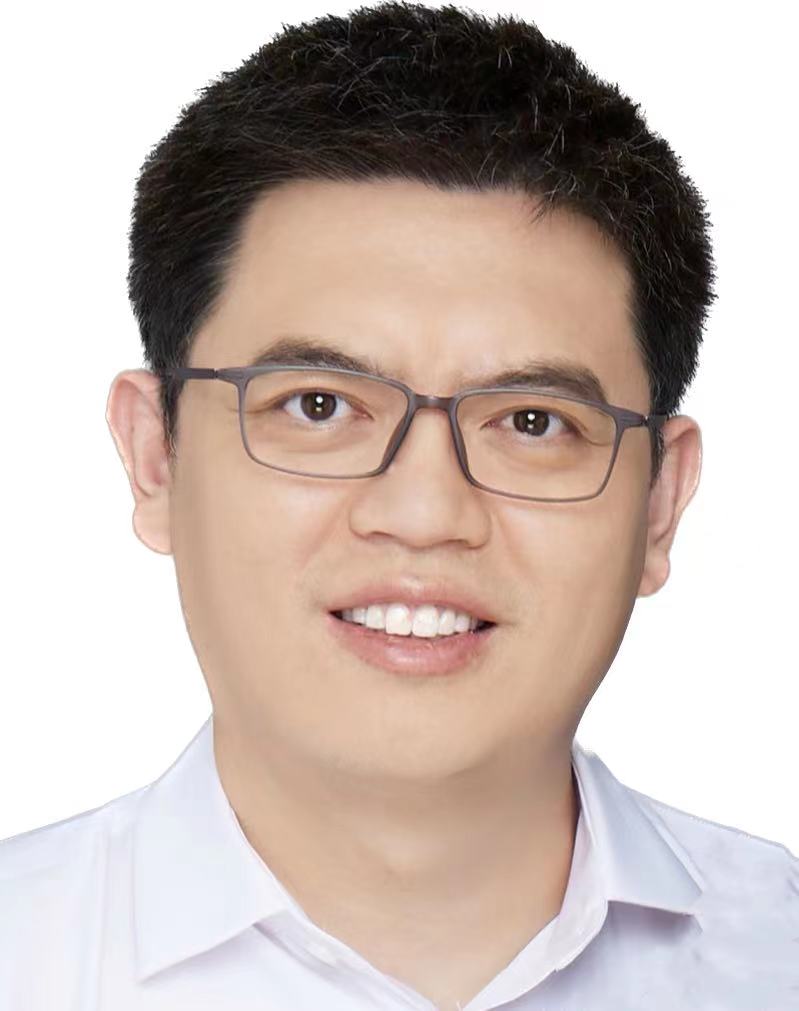}}]{Long Ye}
	received the B.E. degree in Electronic Engineering from Shandong University, Jinan, China, in 2003, and the M.S. and Ph.D. degrees from the Communication University of China, Beijing, China, in 2006 and 2012, respectively. He is currently a Professor at the State Key Laboratory of Media Convergence and Communication, Communication University of China, Beijing, China. Before that, he was a Visiting Scholar at Ryerson University, Toronto. Canada. His research interests include computer vision, image compression, and virtual reality.
\end{IEEEbiography}

\begin{IEEEbiography}
	[{\includegraphics[width=1in,height=1.25in,clip,keepaspectratio]{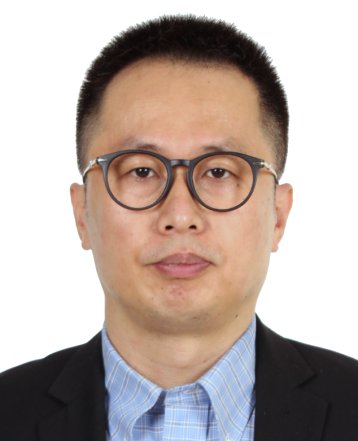}}]{Yi Sun}
	received a master's degree from Peking University in 2011. He is now a PhD student at the School of Cyberspace Science and Technology, Beijing Institute of Technology. In 2022, he was a visiting scholar at the Singapore University of Technology and Design (SUTD). His research interests include Deepfake Detection, Adversarial Examples, and Computer Vision.

\end{IEEEbiography}

\end{document}